\documentclass[a4paper,fleqn,usenatbib]{mnras}

\usepackage[T1]{fontenc}
\usepackage{ae,aecompl}

\usepackage{graphicx}	% Including figure files
\usepackage{amsmath}	% Advanced maths commands
\usepackage{amssymb}	% Extra maths symbols
\usepackage{comment}
\usepackage{color}

\title[]{Weighing the IMBH candidate CO-0.40-0.22* in the Galactic Centre}

\author[A. Ballone et al.]{A. Ballone$^{1}$, M. Mapelli$^{1,2,3}$, M. Pasquato$^{1}$
\\
$^{1}$INAF, Osservatorio Astronomico di Padova, vicolo dell'Osservatorio 5, I-35122 Padova, Italy\\
$^{2}$Institute f\"ur Astro- und Teilchen Physik, Universit\"at Innsbruck, Technikerstrasse 25/8, A-6020 Innsbruck, Austria\\
$^{3}$INFN, Milano Bicocca, Piazza della Scienza 3, I-20126 Milano, Italy}

\date{Accepted XXX. Received YYY; in original form ZZZ}

\pubyear{2018}

\begin{document}
\label{firstpage}
\pagerange{\pageref{firstpage}--\pageref{lastpage}}
\maketitle

\begin{abstract}
The high velocity gradient observed in the compact cloud CO-0.40-0.22, at a projected distance of 60 pc from the centre of the Milky Way, has led its discoverers to identify the closeby mm continuum emitter, CO-0.40-0.22*, with an intermediate mass black hole (IMBH) candidate. We describe the interaction between CO-0.40-0.22 and the IMBH, by means of a simple analytical model and of hydrodynamical simulations. Through such calculation, we obtain a lower limit to the mass of CO-0.40-0.22* of few $10^4 \times  \; M_{\odot}$. This result tends to exclude the formation of such massive black hole in the proximity of the Galactic Centre. On the other hand, CO-0.40-0.22* might have been brought to such distances in cosmological timescales, if it was born in a dark matter halo or globular cluster around the Milky Way.
\end{abstract}


%%%%%%%%%%%%%%%%%%%%%%%%%%%%%%%%%%%%%%%%%%%%%%%%%%

%%%%%%%%%%%%%%%%% BODY OF PAPER %%%%%%%%%%%%%%%%%%

\section{Introduction}\label{intro}

Intermediate-mass black holes (IMBHs), with masses $M_{BH}=10^2\div 10^5 \; M_{\odot}$, represent an ``hollow'' in the mass distribution of detected black holes. Yet, they might even be the missing link between stellar mass and supermassive black holes. 

The LIGO and VIRGO interferometers have indeed first detected the formation of dark objects at the very low end of the IMBHs mass range, by merging of binary stellar mass black holes \citep{Abbott16a,Abbott16b,Abbott17a,Abbott17b,Abbott17c}. 
Indeed, IMBHs might be forming by runaway merging of massive stars in the very dense centre of young massive star clusters \citep[e.g.,][]{Colgate67, Ebisuzaki01, PortegiesZwart04,Freitag06, Giersz15, Mapelli16}, by runaway tidal capture of stars by stellar mass black holes \citep{Miller12,Stone17}, through repeated mergers of stellar black holes in globular clusters \citep{Miller02, Giersz15} or by accretion of stars and compact objects in the disks of active galactic nuclei \citep{McKernan12}. Other possible origins involve the direct collapse of high-mass Population III stars \citep[e.g.,][]{Fryer01,Madau01,Schneider02, Spera17} or collapse of pristine, metal-free gas in high-redshift dark matter halos \citep[e.g.,][]{Bromm03,Begelman06, Lodato07, Shang10, Agarwal12}.

Several possible additional hints of the existence of IMBHs come from the study of kinematics of the central stars and millisecond pulsars of globular clusters \citep[e.g.,][]{Gebhardt02, Gebhardt05, Noyola08, Ibata09, Noyola10,Lutzgendorf11, Lutzgendorf13, FeldmeierKrause13, Lutzgendorf15, Askar17, Kiziltan17, Perera17}, but these are often disputed \citep[e.g.,][]{Baumgardt03,Anderson10, Lanzoni13, Zocchi17,Gieles18}.

IMBHs are also invoked to explain a fraction of the ultraluminous x-ray sources \citep[ULXs; see, e.g.,][]{Miller03, Casella08, Godet09,Sutton12,Mezcua15}. Among these, probably the best candidate is HLX-1 in the S0 galaxy ESO 243-49 \citep{Farrell09}, with black hole mass greater than 500 $M_{\odot}$. Nonetheless, also in the case of ULXs, there is room for alternative theoretical interpretations \citep[e.g.,][]{Goad06, Feng07, Gladstone09, Zampieri09, Mapelli09, Mapelli10, Feng11, Liu13, King14}.

Detections of the upper end of the IMBH mass distribution have also been claimed for dwarf galaxies \citep[see][and references therein]{Kormendy13,Reines15,Mezcua17}.

\citet{Tanaka14}, \citet{Oka16} and \citet{Oka17} have reported the detection, by means of molecular emission lines, of a high velocity compact cloud, CO-0.40-0.22, with internal velocity gradient of several tens of $\mathrm{km \; s^{-1}}$. In particular, through high resolution ALMA observations, \citet{Oka17} set its size, $\lesssim 1$ pc, and its internal line of sight velocity, possibly ranging from -80 to -40 $\mathrm{km \; s^{-1}}$ with respect to us. In addition, these authors discovered an unresolved object, CO-0.40-0.22*, very close to CO-0.40-0.22 and emitting at 231 and 266 GHz in continuum emission. This has been interpreted as an IMBH, with an extremely low accretion rate \citep[but see][for an alternative interpretation]{Ravi18}. \citet{Oka16} and \citet{Oka17} also showed, by means of pure gravity simulations, that the high internal velocity dispersion of the observed cloud can be qualitatively explained by the interaction with a $10^5 \; M_{\odot}$ IMBH. Other physical explanations are possible \citep{Tanaka14, Ravi18, Yalinewich17, Tanaka18} and \citet{Tanaka18} even disputed the correctness of the reduction of the ALMA data by \citet{Oka17} \citep[even though a velocity gradient of 20-40 $\mathrm{km s^{-1}}$ has been confirmed; see Fig. 11 in][]{Tanaka18}.

Nonetheless, even if the result is still highly debated, the possibility that CO-0.40-0.22* is associated with an IMBH makes this object extremely intriguing, because this could be one of the closest IMBHs to us and because of its vicinity to SgrA*. Thus, the aim of this paper is to gain insight on the possible mass of CO-0.40-0.22*, in the IMBH scenario. In particular we built a simple analytical model to describe the interaction between the observed cloud and this putative IMBH, that we also tested through hydrosimulations. This allows to put, for the first time, a lower limit to the mass of this object, that might help understanding its possible nature and origin.

\section{Analytical model}\label{anmod}

As already mentioned, \citet{Oka16} and \citet{Oka17} speculate that the high velocity gradient observed in the cloud CO-0.40-0.22 is generated by the interaction between this cloud and a putative IMBH, located on the position of CO-0.40-0.22*, a closeby mm \citep[and possibly IR,][]{Ravi18} continuum emitter. In \citet{Oka17}, these authors provided a single N-body simulation to try to fit their observations, with IMBH mass of $10^5 \; M_{\odot}$. In this section we try to present a simple analytical calculation to model this interaction; its main purpose is to identify a lower limit for the mass of CO-0.40-0.22*.

\begin{figure}
\begin{center}
\includegraphics[scale=0.36,trim={1cm 2cm 0 0},clip]{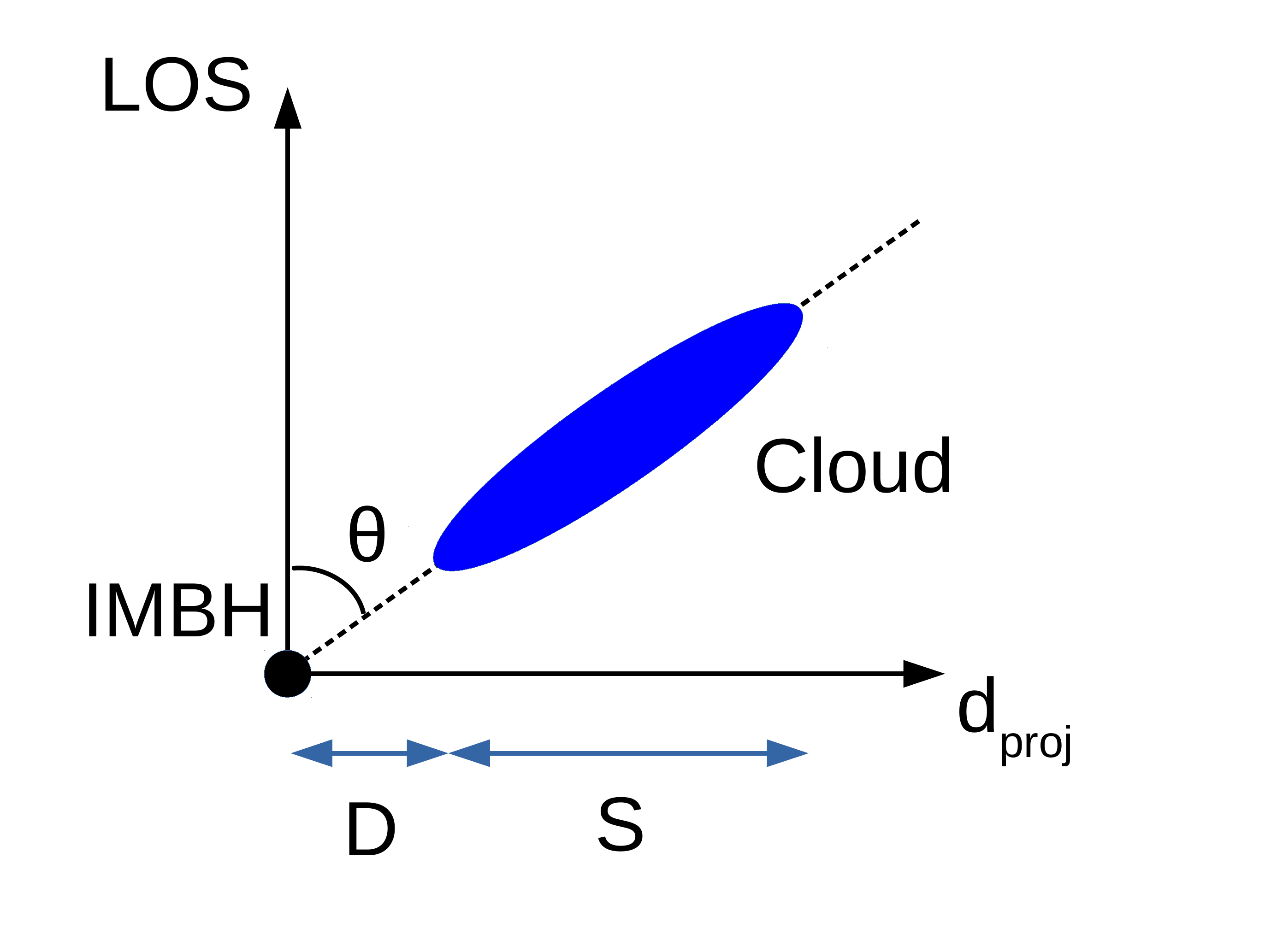}
\caption{Sketch representing the current cloud configuration, in our analytical model. $D$ and $S$ are the projected distance from the IMBH and size of the cloud (see text). 
}\label{sketch}
\end{center}
\end{figure}

In our simple model, we assume that the cloud is radially infalling towards the IMBH. Fig. \ref{sketch} shows a sketch of this configuration. 
Such an assumption is based on few considerations:
\begin{enumerate}
\item The observed position-velocity diagrams show evidence of an ``oblique'' occupation of the cloud in this phase-space. In this case, \textbf{the internal large velocity gradient is to be attributed to a different bulk velocity between different parts of the cloud}, rather than to an homogeneous internal velocity dispersion.
\item \textbf{The cloud and the putative IMBH CO-0.40-0.22* are clearly separated} from each other, both in the ALMA emission map and in the corresponding position-velocity diagrams. Hence, the cloud is not currently experiencing its pericentre passage around the IMBH and the velocity gradient through its elongation can only be explained by its tidal elongation towards the IMBH.
\item \textbf{The highest tidal velocity gradient}, for a fixed IMBH mass, \textbf{is obtained in the case of a radial cloud trajectory}. In other words, a radial trajectory is the one providing the lowest IMBH mass needed to produce the observed velocity gradient.
\end{enumerate}

The ``orbital'' velocity of a radially infalling object is

\begin{equation}\label{vff}
v_{orb}=\sqrt{\frac{2GM_{BH}}{d_{BH}}+2E_{orb}},
\end{equation}

where $G$ is the gravitational constant, $M_{BH}$ is the mass of the IMBH, $d_{BH}$ is the distance from the IMBH and $E_{orb}$ is the specific orbital energy of the cloud. 

$E_{orb}$ can be positive, negative or null, leading to velocities along the orbit that are bigger, smaller or equal to the escape velocity. At the observed distance from 
CO-0.40-0.22* it is fair to assume that the IMBH is the main responsible of the velocity difference between different parts of the cloud and we restrict ourselves to the case of $E_{orb}=0$. Nonetheless, in section \ref{ingred} and in Appendix \ref{eorb} we will discuss the impact of a non-null $E_{orb}$ on our estimate of the mass of CO-0.40-0.22*.

\begin{table*}
\caption{Lower limits for the mass of the IMBH CO-0.40-0.22*}
\label{mbh}
\centering  
\begin{tabular}{l l l l}
\hline
 & $\Delta v_{LOS}=20 \;km s^{-1}$ & $\Delta v_{LOS}=30 \;km s^{-1}$ & $\Delta v_{LOS}=40 \;km s^{-1}$ \\

\hline
$(D,S)=(1,15)$ arcsec & $9.0\times 10^3 M_{\odot}$ & $2.0 \times 10^4 M_{\odot}$ & $3.6\times 10^4 M_{\odot}$ \\ 
$(D,S)=(2,14)$ arcsec & $2.6\times 10^4 M_{\odot}$ & $5.7 \times 10^4 M_{\odot}$ & $1.0 \times 10^5 M_{\odot}$ \\ 
$(D,S)=(3,13)$ arcsec & $5.5\times 10^4 M_{\odot}$ & $1.2\times 10^5 M_{\odot}$ & $2.2\times 10^5 M_{\odot}$ \\ 
\hline
\end{tabular}
\end{table*}

So, in this case, the observed velocity gradient between the head and the tail of the cloud is

\begin{equation}\label{dvlos}
\begin{split}
\Delta v_{LOS} & =  \sqrt{2GM_{BH}}\left[\frac{1}{\sqrt{d_{BH,h}}}-\frac{1}{\sqrt{d_{BH,t}}}\right]\cos\theta \\
& = \sqrt{2G M_{BH} \sin\theta}\cos\theta\left[\frac{1}{\sqrt{D}}-\frac{1}{\sqrt{S+D}}\right],
\end{split}
\end{equation}

where $D$ is the observed (projected) distance between the closest point of the cloud and the IMBH, $S$ the observed (projected) size of the cloud, $\theta$ is the angle between the cloud trajectory and the line of sight (LOS), while $d_{BH,h}$ and $d_{BH,t}$ are the intrinsic distance between the IMBH and the head and the tail of the cloud, respectively (see Fig. \ref{sketch}).

From Eq. \ref{dvlos} we can directly infer that 

\begin{equation}\label{lowlim}
M_{BH}>\frac{3\sqrt{3}}{4}\frac{(\Delta v_{LOS})^2}{G}\frac{D(S+D)}{(\sqrt{S+D}-\sqrt{D})^2},
\end{equation}

with $2/(3\sqrt{3})=\max(\sin\theta\cos^2\theta)$. This corresponds to $\theta \simeq 35^{\circ}$.

The values of $S$, $D$ and $\Delta v_{LOS}$, from the observed position-velocity diagrams in \citet{Oka17}, come with an uncertainty given by the instrumental spread function and by the interpretation of the actual limits of the cloud. We chose optmistic, pessimistic and intermediate contours for the cloud; thus, we calculated the lower limit for $M_{BH}$ for $(D,S)$=$(1,15)$,$(2,14)$,$(3,13)$ arcsec \footnote{22 arcsec = 0.9 pc \citep{Oka17}.} and $\Delta v_{LOS}$=20,30,40 $\mathrm{km\; s^{-1}}$. The results are summarized in Table \ref{mbh}.
We stress that the value of $M_{BH}$ obtained with Eq. \ref{lowlim} is already the smallest possible mass for the putative IMBH CO-0.40-0.22*, due to our extreme assumption of a radial orbit for the gas cloud CO-0.40-0.22.

\subsection{Impact of other physical ingredients}\label{ingred}

As already mentioned in section \ref{anmod}, we restricted our calculation to $E_{orb}=0$, which is the easiest case and a relatively good approximation at those distances 
from the IMBH. 

In the context of a radial orbit, $E_{orb}>0$ corresponds to a non-null velocity at infinity. This is certainly possible: e.g., stars might have velocity dispersion of up to 100 $\mathrm{km \;s^{-1}}$ at those distances from SgrA* \citep[e.g.,][]{Fritz16}. However, this case would further increase the IMBH mass needed to perturb the cloud (see also Appendix \ref{eorb}), hence it is not changing our results, that consist of lower limits.

On the other hand, $E_{orb}<0$ is equivalent to a cloud starting its plunge towards the IMBH with zero velocity at a finite distance. This might happen if the cloud was previously strongly scattered by some other close-by massive object, being left with zero velocity, relative to the IMBH. However, in order to really affect the lower limit on the mass of the IMBH, this should happen at distances comparable with the current position of the cloud. Even assuming that the cloud started its plunge at the current position of its tail, our result on the limit for the mass of CO-0.40-0.22* would be reduced by around a factor of 2 (see Appendix \ref{eorb}).

In the present calculation we neglect the possible effect of the self-gravity of the cloud. Such assumption is based on the fact that self-gravity would oppose the tidal stretching by the IMBH, hence requiring a even higher mass for CO-0.40-0.22*. Nonetheless, the cloud size and its distance from the putative IMBH are of the same order. In this case, from a simple ``Roche limit'' argument, the self-gravity of the cloud would become important only if the cloud mass were comparable to the mass of the IMBH, which is not realistic. 

We also neglect the impact on the cloud of the tidal force of the central supermassive black hole (SMBH) of the Milky Way \citep[$M_{SMBH}\simeq 4 \times 10^6 \; M_{\odot}$;][]{Boehle16, Gillessen17} and of the Galactic Centre stars \citep[$M_*(<60\; \mathrm{pc})\simeq 10^8\; M_{\odot}$;][]{Launhardt02,Fritz16,FeldmeierKrause17,GallegoCano18}. In fact, applying Eq. \ref{dvlos}, the velocity gradient produced on CO-0.40-0.22 by the SMBH and the stellar component at a distance of 60 pc from the Galactic Centre would approximately be 0.5 and 3 $\mathrm{km \;s^{-1}}$, respectively. Both values are significantly smaller than the observed velocity gradient.

Thermodynamics, turbulence and magnetic fields can be important in typical molecular clouds. However, in this case, the thermal, turbulent and magnetic pressures in the cloud should be comparable to the tidal "pressure" of the IMBH.  

In particular, concerning thermodynamics, we can exclude that the observed velocity gradient is produced by a strong unbalance between the internal cloud pressure and the external pressure of the surrounding interstellar medium. In fact, such a high velocity gradient would mean either an ``explosion'' or an ``implosion'' of the cloud and requires another physical process \citep[such as those invoked by][]{Tanaka14,Tanaka18,Ravi18, Yalinewich17} to explain such an out-of-equilibrium configuration. However, we can easily test whether internal pressure gradients induced by radiation and hydrodynamical cooling/heating can become as important as tides. Hence, we compare the acceleration driven by a pressure difference $\Delta P$ over the cloud size $S/\sin \theta$,

\begin{equation}
a_{th}=\frac{1}{\rho_{cl}}\frac{\Delta P}{S}\sin\theta
\end{equation}

to the tidal acceleration\footnote{We must note that this is a first order approximation, not fully correct at the observed distance between the IMBH and the cloud, but still useful for such a back-of-the-envelope estimate.}

\begin{equation}
a_{tid}=\frac{S}{\sin\theta} \frac{G M_{BH}}{d^3_{BH}}.
\end{equation}

In order for these to be comparable,

\begin{equation}
\frac{\Delta P}{P} = \left(\frac{S}{\sin\theta}\right)^2\frac{ GM_{BH}}{d^3_{BH}}\frac{\mu m_H}{k_B T_{cl}},
\end{equation}

where $\mu$ is the mean molecular weight, $m_H$ is the hydrogen mass and $k_B$ is the Boltzmann constant.
For the case $(D,S)$ = (2,14) arcsec and $\Delta v=30\; \mathrm{km \; s^{-1}}$ and assuming $d_{BH}=(S+D)/(2\sin\theta)$, $\theta=35^{\circ}$, $\mu=2.46$ and $T_{cl}=60 \;K$, we get $\Delta P/P\simeq 5 \times 10^3$. This ratio means that, in order for thermodynamical evolution to matter, the cloud should get extremely compressed, which is not happening for pure tidal evolution, or get heated to temperature bigger than $10^4$ K, meaning a change of status to atomic or even ionized state. The gas would then not shine in molecular lines at all.

We can also try to estimate whether we expect turbulence to have an impact on the cloud, comparable to that of the tidal field. In fact, a cloud of 40 $M_{\odot}$ in virial equilibrium requires a turbulent $\sigma_v\approx\sqrt{GM_{cl}/R_{cl}}\approx\sqrt{2GM_{cl}\sin\theta/S} \approx 0.5\; \mathrm{km \; s^{-1}}$, for S=14 arcsec. In our model, the observed velocity gradient (20-40 $\mathrm{km \; s^{-1}}$) is produced by the tidal stretching of the IMBH. In order for turbulence to compete with the tidal stretching, the cloud should be in an extremely supervirial configuration, i.e., we would require, again, some other physical phenomenon to explain such an out of equilibrium state.

To test the impact of magnetic fields, we can use the magnetic stability criterion derived by \citet{Mouschovias76}. In this case, an uniform magnetic field B is able to support the cloud against gravitational collapse if

\begin{equation}\label{magcoll}
B\geq\frac{M_{cl}}{73\; M_{\odot}} \left(\frac{R_{cl}}{0.1 \;\mathrm{pc}}\right)^{-2} \mathrm{mG}
\end{equation}

We also mentioned before that the Roche mass for the cloud should be of the order of the mass of the IMBH. Hence, if we use $M_{cl}=M_{BH}$ in Eq. \ref{magcoll}, we can also get an estimate of the magnitude of the magnetic field needed to have an effect comparable to that of the tidal force. If we approximate again $R_{cl}$ in Eq. \ref{magcoll} with S/(2$\sin\theta$) and consider the case with $(D,S)$ = (2,14) arcsec and $\Delta v=30\; \mathrm{km \; s^{-1}}$, we get $B\gtrsim 40$ mG, which is much higher than the typical magnetic field in molecular clouds, even in the Galactic Centre \citep[where it is estimated to be maximum few mG; e.g.,][]{Ferriere09}. Furthermore, given the 
estimated cloud mass, $M_{cl}=40 M_{\odot}$, we rather expect a a magnetic field of the order of 30 $\mu$G, if it were in a critical state.

\section{Simulations}

\subsection{Initial conditions and methods}\label{ic}

In order to test our analytical calculation, we run an hydrodynamical simulation with the Eulerian Adaptive Mesh Refinement (AMR) code RAMSES \citep{Teyssier02}. We adopted Cartesian coordinates and chose a cubic box with $x,y,z=[-5.5:5.5]$ pc. 

The IMBH is initially put at $(x,y,z)=0$ as a sink particle, with Bondi-Hoyle accretion \citep{Bleuler14}. Even though no significant motion of the IMBH is expected (given the small cloud mass), we allow it to move and integrate its motion by direct force summation.

For the cloud, we chose the simplest case of a radial trajectory with $E_{orb}=0$. In this case, the time needed to reach the IMBH from a certain distance is 

\begin{equation}\label{tfall}
t_{fall}=\sqrt{\frac{2d^3_{BH}}{9GM_{BH}}},
\end{equation}

whenever the mass of the IMBH is significantly larger than the mass of the cloud (which is the case of our model, see section. \ref{anmod}). Eq. \ref{tfall} then allowed us to set the initial conditions of our simulation. Specifically, we can get the initial distance of head and tail of the cloud as 

\begin{equation}\label{distini}
d_{BH,0}=4.5 G M_{BH} (t_{fall}+t_{arb})^{1/3},
\end{equation}

where $t_{arb}$ is any arbitrary time, needed to head and tail of the cloud to reach their current position. 

In particular, we tested the case with $(D,S)=(2,14)$ arcsec and $\Delta v_{LOS}=30\; \mathrm{ km \;s^{-1}}$, with $\theta \simeq 35^{\circ}$ and corresponding $M_{BH}= 5.7 \times 10^4 M_{\odot}$ (see Table \ref{mbh}). For this model, $d_{BH,h}\simeq 0.14$ pc and $d_{BH,t}\simeq 0.99$ pc. We adopted $t_{arb}=0.3$ Myr, which gives $d_{BH,h,0}\simeq 4.73$ pc and $d_{BH,t,0}\simeq 5.01$ pc. We hence put a spherical cloud on the x-axis, with radius $R_{cl}=(d_{BH,t,0}-d_{BH,h,0})/2\simeq 0.14$ pc at a distance of $d_{BH,cl,0}=(d_{BH,h,0}+d_{BH,t,0})/2\simeq 4.87$ pc from the IMBH and with a velocity towards the IMBH of $\simeq 10\;\mathrm{ km \;s^{-1}}$ (see Eq. \ref{vff}). We adopted a cloud mass $M_{cl} = 40 M_{\odot}$ (which corresponds to a density of $\rho_{cl}\simeq 2.3 \times 10^{-19} \;\mathrm{g \; cm^{-3}}$), a cloud temperature of $T_{cl}=60$ K (see section \ref{anmod}) and put it in pressure equilibrium with a rarified and hot background medium with $\rho_{bg}=\rho_{cl}/10^5$.

Compared to the test-particle and N-body simulations in \citet{Oka16} and \citet{Oka17}, we considered completely different initial conditions for the cloud. In fact, as discussed in section \ref{anmod}, we chose the value of the mass estimated by these authors from its emission, which leads to a cloud that is not bound by its self-gravity and does not require strong turbulence to support it. This value is much smaller than that of their simulated cloud, with $M_{sim}=10^3 M_{\odot}$. In particular, in \citet{Oka17} the authors distribute their particles with a Gaussian radial mass distribution with dispersion of $\sigma_r = 0.2$ pc and internal velocity dispersion of $\sigma_v = 1.43 \mathrm{km \; s^{-1}}$, claiming that this leads to an initially virialized cloud. However, in this case, $GM_{sim}/\sigma_r \approx 10 \sigma^2_v$, hence this setup is actually strongly subvirial.\footnote{In fact, calculating the free-fall time $t_{ff}$ of such a cloud and comparing it to the turbulence crossing time $t_{cross}$ and to the time $t_{ca}$ that it takes for their cloud to reach the closest approach to the IMBH, we get $t_{ff}\approx t_{cross}/2 \approx t_{ca}/10$. Hence, the final result of the N-body simulations presented in \citet{Oka17} might suffer of a strong imprint of these unstable initial conditions.} A virial equilibrium would require a $\sigma_v \approx 5 \mathrm{km \; s^{-1}}$. If the molecular gas has a temperature of 60 K, as estimated by \citet{Oka17}, this means that the turbulence has a Mach number $\mathcal{M}\approx 10$, which is probably too high for clouds with that size \citep[e.g.,][]{Larson81}.

So, as discussed here and in Section 2.1, we do not expect self-gravity and turbulence to significantly affect the results of our analytical calculation. Nonetheless, we decided to test the influence of self-gravity and turbulence by running two simulations of the same cloud: in the first simulation, the cloud is assumed to have no initial turbulence and no self-gravity, while in the second setup we include gas self-gravity and turbulence.

In particular, for the turbulent cloud we generated a random Gaussian,
divergence-free turbulent velocity field, with power spectrum
$||\delta{}^2_v  || \propto{} k^{-4}$ . Such power spectrum is usually chosen to reproduce the observed trend of the velocity dispersion, in molecular clouds, with the cloud size and the size of its subregions
(Larson 1981). The ratio between kinetic and gravitational
energy is set to 1, i.e., the cloud is marginally self-bound.

The minimum and maximum refinement levels in our simulations are 4 and 10, respectively, which correspond to $(\Delta x,y,z)_{max}= 11/2^4\;\mathrm{pc}=0.6875$ pc and $(\Delta x,y,z)_{min}= 11/2^{10}\; \mathrm{pc}\simeq 0.0107$ pc, respectively. The AMR strategy we used is ``quasi-Lagrangian'', i.e., we refine every cell with mass higher than a certain value $m_{res}$. We chose $m_{res}\simeq 9\times 10^{-4} \; M_{\odot}$, so to be sure to have maximum resolution over the whole cloud, while keeping the backgroung medium at low resolutions.

This test simulation is run with an isothermal equation of state for the gas. Such simplification is justified by the fact that at the cloud densities, the molecular gas is kept at a roughly constant ``equilbrium'' temperature by heating and cooling processes \citep[e.g.,][]{Larson85,Larson05,Koyama00}. 

In terms of numerics, we adopted an ``exact'' Riemann solver with 10 iterations and a MonCen slope limiter for the piecewise linear reconstruction \citep[e.g.,][]{Toro09}.

\begin{figure}
\begin{center}
\includegraphics[scale=0.82]{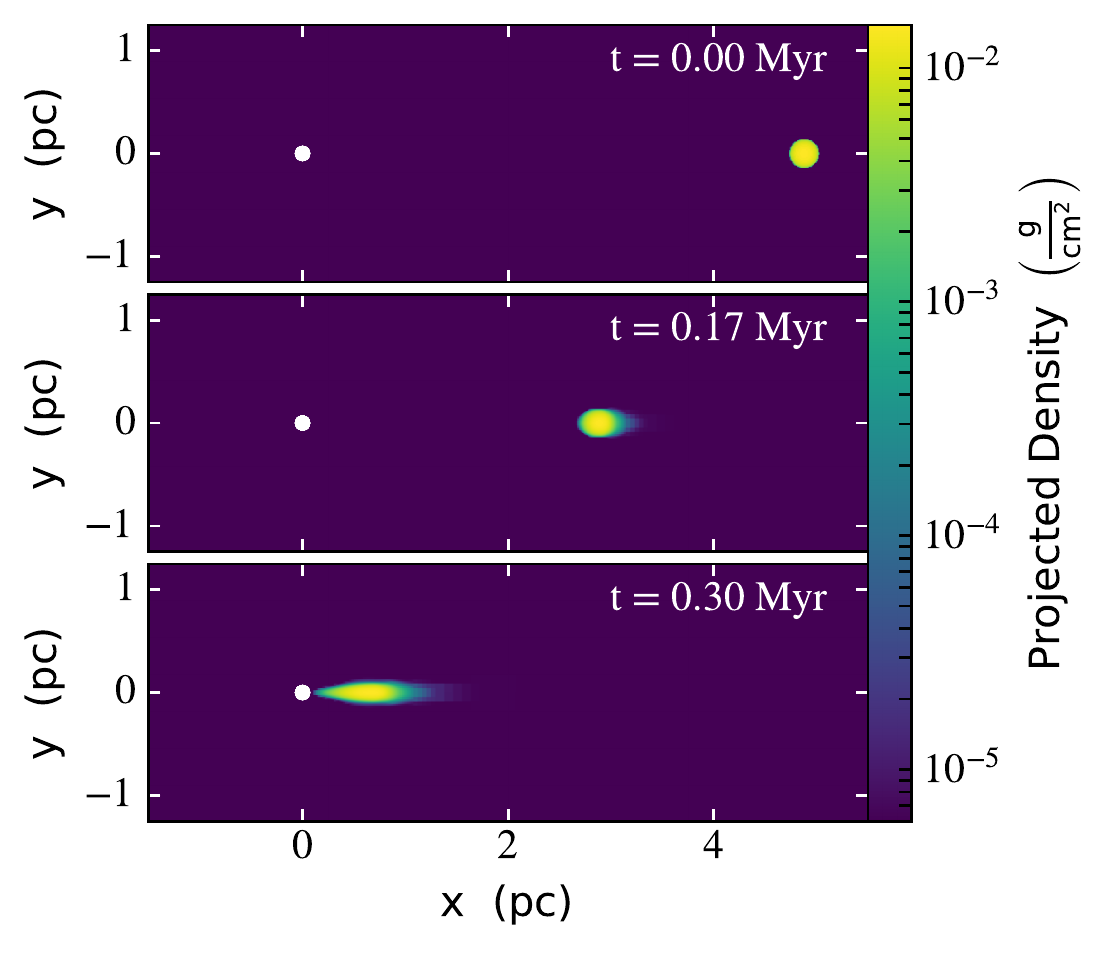}
\caption{Projected density maps for the run with no self-gravity and no turbulence. The figure shows the evolution of the cloud on its radial orbit at $t=$ 0, 0.17 and 0.30 Myr, in the simulation domain. The white dot shows the position of the IMBH.  
}\label{intmap}
\end{center}
\end{figure}

\begin{figure*}
\begin{center}
\includegraphics[scale=0.45]{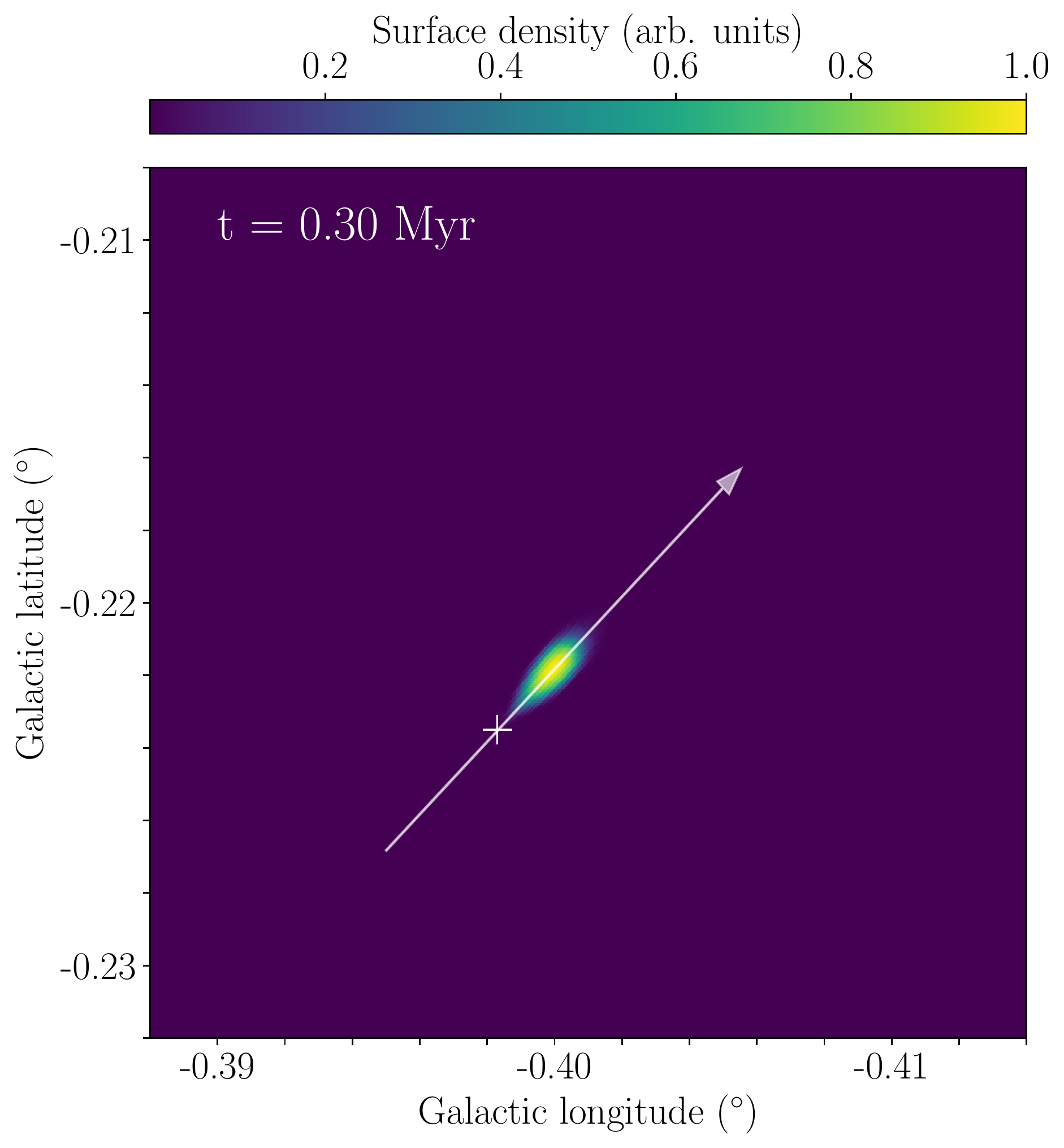}
\includegraphics[scale=0.45]{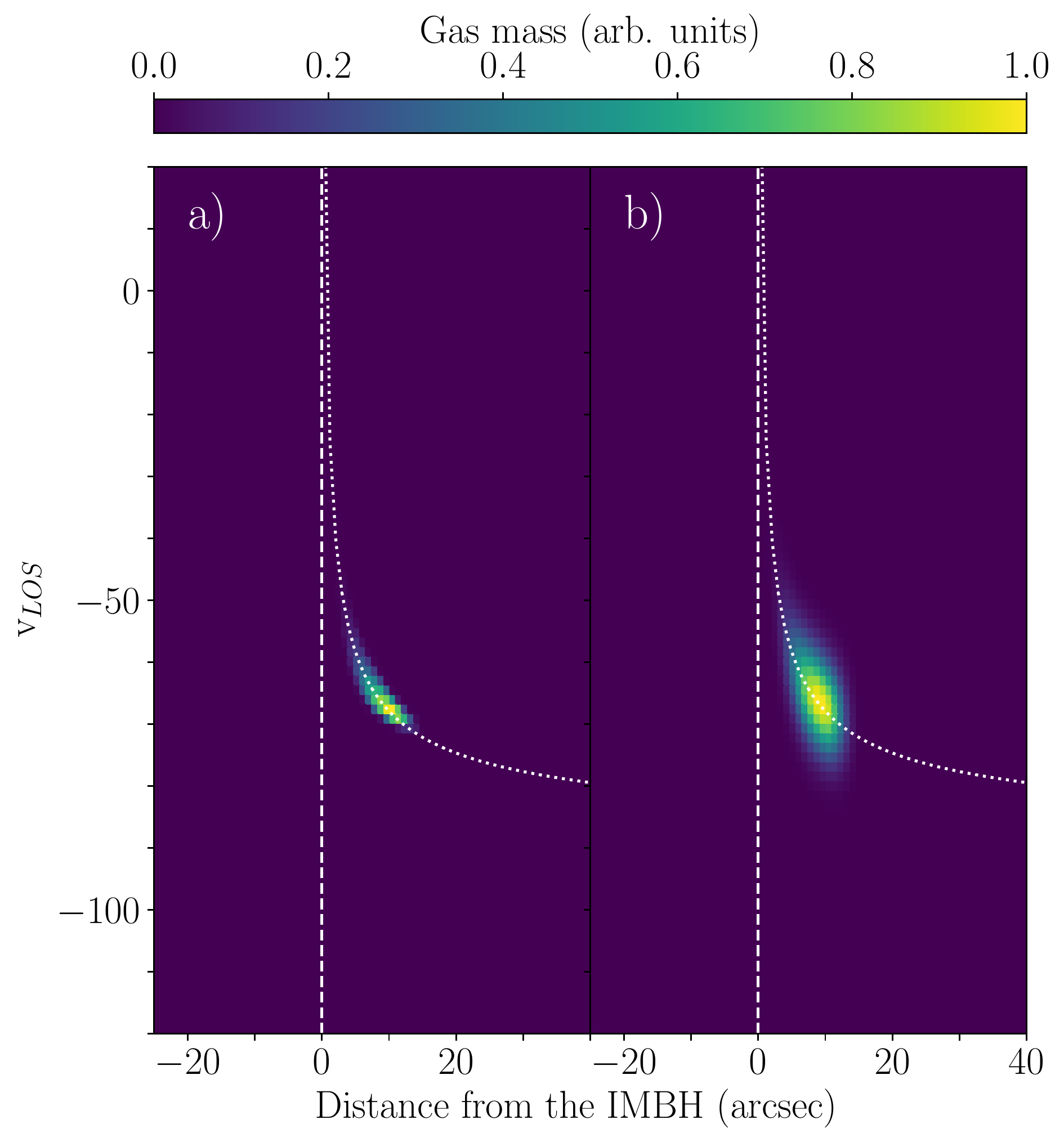}
\caption{Run with no self-gravity and no turbulence. Left: surface density maps of the simulated cloud on the sky plane (see text), at $t=0.3$ Myr. The arrow shows the direction of the simulation's x-axis, i.e., the axis along which the position-velocity diagrams are calculated. Right: position-velocity diagrams for the simulated cloud. The vertical dashed line marks the position of the IMBH, the dotted line represents the analytical solution for the line-of-sight velocity of a point-mass on a parabolic orbit. We subtracted a velocity of $-90 \; \mathrm{km \; s^{-1}}$ to all velocities, to match the observations (see text). Panel (b) is obtained from panel (a) after applying a gaussian smoothing with FWHM equal to 1.2 arcsec and 11 km/s in distance from the IMBH and line-of-sight velocity, respectively. These plots can be qualitatively compared to Fig. 1 and 2 in \citet{Oka17}.
}\label{obspl}
\end{center}
\end{figure*}

\subsection{Results}

In Fig. \ref{intmap} we show the free-fall of the cloud towards the IMBH in our simulation. As the cloud approaches its attractor, the tidal force progressively distorts its shape, leading to an elongation in the direction of the motion and a perpendicular compression. In the lower panel of Fig. \ref{intmap} the distance from the IMBH of the head of the cloud is much smaller than the cloud size. For this reason, the simplified treatment of tidal effects based on a MacLaurin expansion of the gravitational force is no longer valid. The latter would have predicted a tidal force acting symmetrically, with respect to the cloud baricentre, onto head and tail of the cloud. This approximation is no longer valid at this time of the simulation and the cloud assumes a drop-like shape, instead. We must stress here that the positions of head and tail of the cloud and, consequently, the cloud elongation are a direct consequence of our imposition of Eq. \ref{distini}, once $t_{arb}$ is chosen. On the other hand, the cloud thickness in the direction perpendicular to the orbit depends on what shape is initially given to the cloud, i.e., on its initial thickness.

\begin{figure*}
\begin{center}
\includegraphics[scale=0.45]{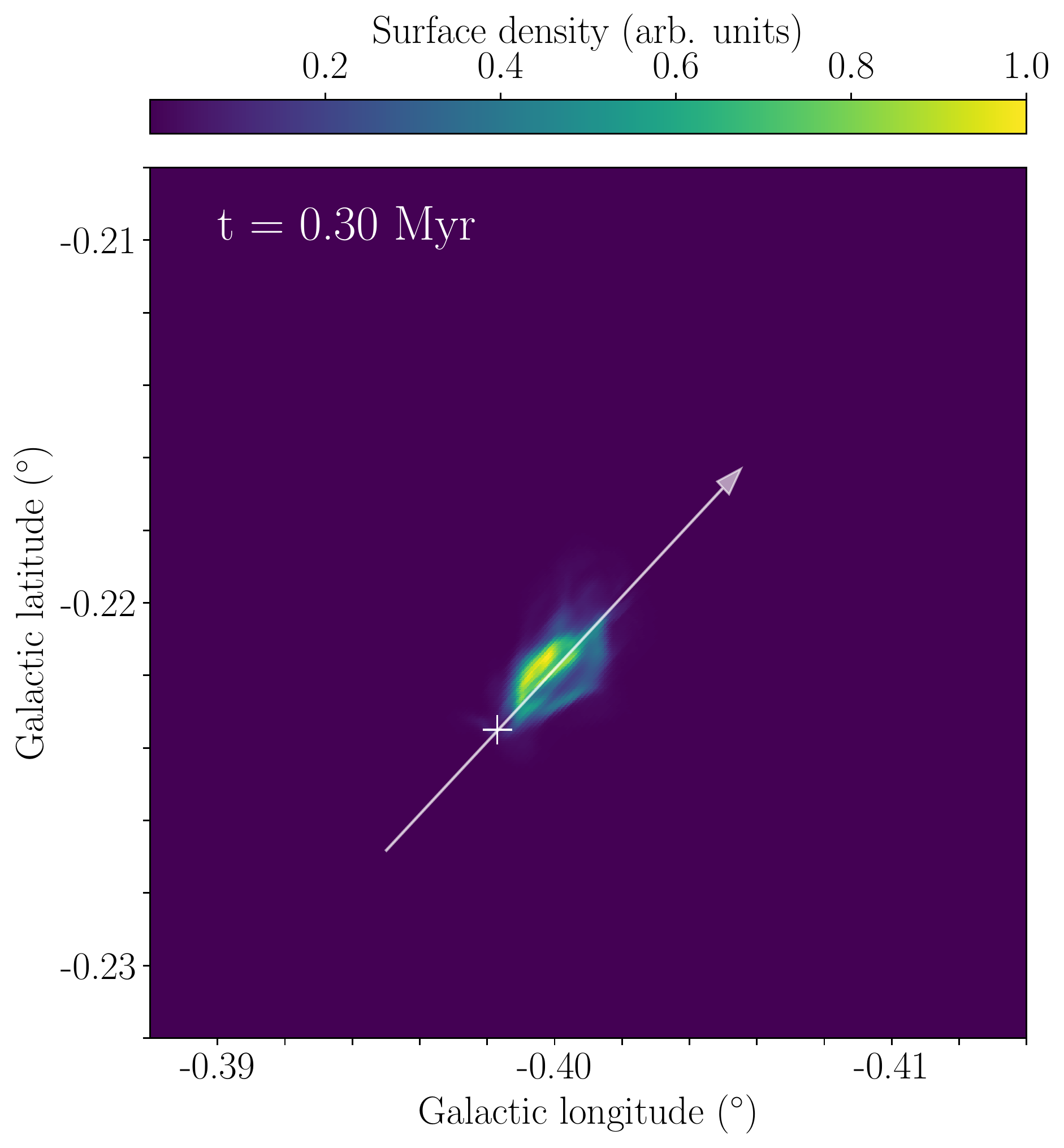}
\includegraphics[scale=0.45]{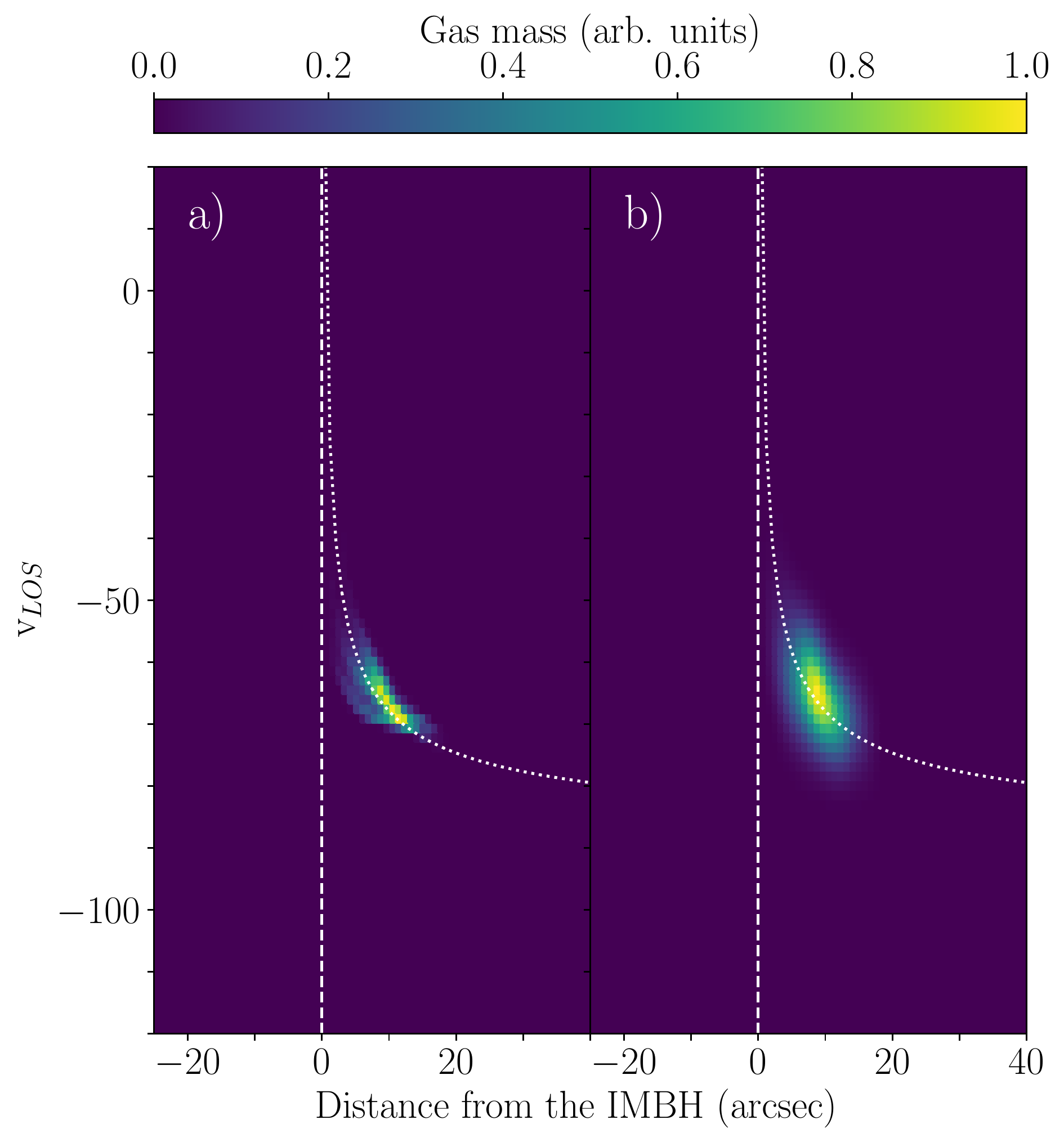}
\caption{Same as Fig. \ref{obspl}, but for the run including self-gravity and turbulence (see text).
}\label{obsplturb}
\end{center}
\end{figure*}

The left panel of Fig. \ref{obspl} shows the cloud as projected to the sky plane. In order to produce this plot, we simply assumed that the line-of-sight forms an angle $\theta \simeq 35^{\circ}$ with the x-axis of our simulation (see Fig. \ref{sketch} and section \ref{ic}) and an angle $\alpha=45^{\circ}$ with the Galactic plane (the arrow shows the direction of the x-axis of the simulation). Interestingly enough, also the observed cloud CO-0.40-0.22 shows a sort of drop-like shape, vaguely similar to the simulated cloud shown in Fig. \ref{obspl}. 

The right panel of Fig. \ref{obspl} shows mock position-velocity diagrams for the simulated cloud. The distance from the IMBH is simply $d_{proj}=d_{BH}\cos\theta=x\cos\theta$, while the line-of-sight velocity in this plot is obtained by adding a fixed velocity of $-90 \; \mathrm{km \; s^{-1}}$ to the gas velocity. The latter should represent the velocity of the centre of mass of the whole IMBH+cloud system and it is needed to match the observed velocity values. Such centre of mass velocity is nonetheless compatible with the typical orbital velocities at those distances from the Galactic centre. However, we must point out that we would obtain a different centre of mass velocity by simply assuming that the cloud is falling towards the IMBH on an orbit with $E_{orb}\neq 0$ in Eq.  \ref{vff}.

For these simple diagrams we plotted the mass in any position-velocity bin, rather than the emission. This is a 0th-order approximation , based on the assumption that every molecular species has an uniform abundance over the whole cloud and that the observed molecular lines are optically thin (which is probably the case, given the cloud properties). Panel (a) is the direct risult of the simulation, while panel (b) is obtained from panel (a) by applying a Gaussian smoothing with with FWHM equal to 1.2 arcsec and 11 $\mathrm{km\; s^{-1}}$ in distance from the IMBH and line-of-sight velocity, respectively. Applying a Gaussian smoothing is needed to reproduce the observed velocity dispersion at any fixed distance from the IMBH. Such velocity dispersion might be explained by internal supersonic turbulence. However, a velocity dispersion of 11 $\mathrm{km\; s^{-1}}$ in a gas at 60 K implies a turbulence with Mach number $\mathcal{M}\gtrsim 20$, which seems very unlikely. Thus, the most likely explanation is that the observed velocity dispersion is due to the instrumental spread-function. Indeed, the position-velocity diagrams in Fig. 2 of \citet{Oka17} show large velocity dispersion for all the noise/background patches surrounding CO-0.40-0.22.

The head of the cloud appears more rarefied (i.e., less visible) in the simulated position-velocity diagrams. This is simply because the leading part of the cloud occupies a portion of the orbit with larger velocity gradient (see the dotted line in the right panels of Fig. \ref{obspl}). Hence, its emission (or, for our simplified comparison, its mass) will be spread over more velocity bins, compared to the trailing part. In addition, as discussed before, the cloud assumes a drop-like shape close to the IMBH, thus leading to a non-uniform mass distribution along the length of the cloud. The observations by \citet{Oka17} show a large ($\simeq 20-40\; \mathrm{km \; s^{-1}}$) velocity gradient in the brightest gas and some possible emission from very high (relative) velocity gas. In our model, this would possibly imply an ``effective'' larger velocity gradient between the head and the tail of the cloud and, consequently, an even higher black hole mass. 

Fig. \ref{obsplturb} shows the density map and the position-velocity diagrams of the simulation in which the effects of self-gravity and turbulence have been included. As visible, the turbulence leads to a non-uniform distribution of the gas. This is indeed more consistent with the observed cloud, which shows some subpeaks in its elongation. The main point of this work is still confirmed in the lower panel of Fig. \ref{obsplturb}, where the cloud has basically the same extent in the position-velocity space as in Fig. \ref{obspl}. Nonetheless, the turbulence gives a slightly larger velocity extent close the head of the cloud, compared to our simpler uniform model.

So, including turbulence gives results that are slightly closer to the observations. However, turbulence does not significantly affect the overall velocity gradient produced across the whole cloud length by the tidal field of the IMBH.

\section{Discussion and conclusions}

In this paper, we assumed that the large velocity gradient observed in the very compact molecular cloud CO-0.40-0.22 \citep{Oka16, Oka17} is the result of the infall of this cloud on the putative IMBH CO-0.40-0.22*. Our extreme assumptions (e.g., radial infall, best possible inclination angle between the cloud orbit and the sky plane, etc.) gave us a strong lower limit to the mass of such IMBH of few $\times 10^4 M_{\odot}$. We must again stress that the lower limits in Table \ref{mbh} are obtained assuming the most favourable conditions and higher masses are to be expected. This is the first paper where a robust lower limit is given. However, we cannot exclude that other phenomena explain the observed velocity gradient inside CO-0.40-0.22, such as collisions with other clouds, bipolar outflows from young stellar objects or supernova explosions \citep{Tanaka14, Ravi18, Yalinewich17}. 

Is it reasonable to find such a massive black hole at that distance from the Galactic Centre?

An estimate of its dynamical friction timescale ($t_{df}$) can help us answering this question. In fact, if $t_{df}$ were short, it would be very unlikely to find it at its current position.  At 60 pc from SgrA* (i.e., the projected distance of CO-0.40-0.22*), the IMBH would interact with the stars of the Milky Way nuclear star cluster and the inner parts of the nuclear stellar disc \citep[see][and references therein]{BlandHawthorn16}. At those distances, the enclosed stellar mass can be described by a power-law $M(r)=2\times M_0 (r/R_0)^\alpha$, with $M_0=2 \times 10^8 \; M_{\odot}$, $R_0=60 pc$ and $\alpha=1.2$ \citep[e.g.,][]{Fritz16}. Using Eq. 16 in \citet{McMillan03},

\begin{equation}
t_{df}=\frac{\alpha+1}{\alpha(\alpha+3)}\frac{1}{\chi\ln\Lambda}\left(\frac{M_0}{G}\right)^{1/2}\frac{R^{3/2}_0}{m_{BH}},
\end{equation}
and assuming $\chi=0.34$ (if the IMBH moves at the circular velocity and the stars are in dynamical equilibrium) and $\ln\Lambda=1-10$, we get that $t_{df, BH}\simeq 1.3-13$ Gyr for $m_{BH}=5\times 10^4 \; M_{\odot}$. This means that this IMBH has probably spent quite some time at its current position and it is not expected to get much closer in the next few Gyr. This calculation holds as long as the IMBH has reached its current position in isolation, i.e., if it is currently not surrounded by a host stellar cluster (see later).

Concerning its origin, similar IMBH masses are obtained by theoretical estimates of BHs forming in high-$\sigma$ density fluctuations in a $\Lambda$CDM cosmological context \citep[e.g.,][]{Volonteri03}. By means of N-body
cosmological simulations, \citet{Diemand05} have shown that 10-100 of such IMBHs are expected to be found in the inner kpc of the Milky Way. A similar result has also been obtained by the semi-analytic work by \citet{Volonteri05}, also including dynamical friction of the IMBH host halo. Unfortunately, such studies barely (or do not) reach distances from the centre of the galaxy smaller then 100 pc. 

The lower limit to the IMBH mass found in the present study is, instead, only marginally compatible with (at the upper end of) the IMBH mass distribution of putative IMBHs in globular clusters \citep[e.g.,][]{Lutzgendorf13, Mezcua17} and it is also in tension with theoretical estimates (e.g., \citealp{Miller02, PortegiesZwart04, Freitag06, Mapelli16}; but see also \citealp{Giersz15}).

The infall of globular clusters by dynamical friction has been theorized to be responsible for the formation of nuclear star clusters \citep{Tremaine75, Capuzzo93, Capuzzo08, Agarwal11}, also in the case of the Milky Way \citep{Antonini12,Gnedin14}, and perhaps for the formation/growth of supermassive black holes \citep[e.g.,][]{Ebisuzaki01, PortegiesZwart06}.

In particular, \citet{Mastrobuono14} predicted that IMBHs hosted by globular clusters are expected to inspiral down to the inner pc of the Galaxy. Their simulations, though, assume that massive ($\approx 10^6 M_{\odot}$) globular clusters can survive up to distances of 20 pc from SgrA* \citep[in this regard, see][]{Miocchi06}. 

\citet{PortegiesZwart06} also predicted a population of around 50 IMBHs in the inner 10 pc of the Milky Way. However, these are expected to have masses of the order of $10^3 M_{\odot}$, since they should be born in lower mass ($\approx 10^5 M_{\odot}$) star clusters, forming closer to the Galactic Center, such as the Arches \citep{Nagata95,Cotera96,Serabyn98} and the Quintuplet \citep{Nagata90, Figer95}. Finally, \citet{Fragione18} modeled the fate of IMBHs born in globular clusters in the halo of Milky-Way-like galaxies and found that the most massive ($\gtrsim 10^6 M_{\odot}$) globular clusters might have delivered few massive ($\gtrsim 10^4 M_{\odot}$) IMBHs at distances smaller than 100 pc from the centre of the galaxy. In contrast with the assumptions of \citet{Mastrobuono14}, \citet{McMillan03} and \citet{Fragione18} expect the parent globular cluster to have dissolved before reaching the observed position of CO-0.40-0.22* and have left a ``naked'' black hole. From the IR observations of CO-0.40-0.22 \citep{Ravi18}, it is almost impossible to understand whether the IMBH candidate is naked or surrounded by a star cluster, because of the high absorption along the line of sight.

These arguments show that it is very unlikely that such a massive IMBH formed on the spot: Arches/Quintuplet-like star clusters would not be massive enough to produce IMBHs with mass $>10^4 M_{\odot}$. A more massive local parent cluster, instead, should have been dragged, along with its IMBH, to much smaller distances from SgrA*. Thus, the most plausible scenario is that this object might have formed in the halo of our Galaxy and was successively brought at its current position by
its original host, which dissolved on the way.

Hence, assuming that CO-0.40-0.22* is an IMBH with high mass, as those resulting from our calculation, is not in tension with current theoretical estimates. 

As already mentioned, different explanations for the high velocity gradient of CO-0.40-0.22 are possible. A supernova explosion inside the cloud would provide enough energy to generate it \citep{Tanaka14, Yalinewich17} and it is a feasible alternative. Indeed, \citet{Ravi18} reported that the cloud might be associated to an HII region. On the other hand, as already mentioned by \citet{Oka17}, CO-0.40-0.22 does not clearly show a cavity at its center. A bipolar outflow could be another possibility, but it does not seem to be energetic enough \citep{Tanaka14}. A cloud-cloud collision seems to be the most promising alternative explanation, as discussed by \citet{Tanaka14} and \citet{Tanaka18}. Indeed, CO-0.40-0.22 seem to be on the rim of two large molecular shells. Concerning CO-0.40-0.22*, \citet{Ravi18} showed that its spectral energy distribution can be due to synchrotron emission by an advection dominated accretion flow or by a relativistic jet/outflow (similarly to the case of SgrA*). However, these authors also show that thermal black-body emission from a massive protostellar disc around a young star can be a viable alternative explanation.

In conclusion, the interpretation of the nature of CO-0.40-0.22 and CO-0.40-0.22* is still highly debated, so further attention should be given to this exotic object, particularly in the light of the parallel claim of another IMBH with mass $\gtrsim 10^4 \; M_{\odot}$ in the IRS13E complex at $\approx 0.13$ pc from SgrA* \citep{Schodel05, Fritz10,Tsuboi17}.

\section*{Acknowledgements}

We thank the referee, Prof. Miros\l{}aw Giersz, for his stimulating comments. AB and MM acknowledge financial support from the MERAC Foundation, through grant `The physics of gas and protoplanetary discs in the Galactic centre', and from INAF, through PRIN-SKA `Opening a new era in pulsars and compact objects science with MeerKat'. MP acknowledges support from the European Union's Horizon $2020$ research and innovation programme under the Marie Sk\l{}odowska-Curie grant agreement No. $664931$. AB would like to thank the whole ForDyS group for useful discussions. Most of the simulation post-processing was carried out with the yt toolkit \citep{Turk11}.

%%%%%%%%%%%%%%%%%%%%%%%%%%%%%%%%%%%%%%%%%%%%%%%%%%



\bibliographystyle{mnras}
\bibliography{mylit} % if your bibtex file is called example.bib

\begin{thebibliography}{}
\makeatletter
\relax
\def\mn@urlcharsother{\let\do\@makeother \do\$\do\&\do\#\do\^\do\_\do\%\do\~}
\def\mn@doi{\begingroup\mn@urlcharsother \@ifnextchar [ {\mn@doi@}
  {\mn@doi@[]}}
\def\mn@doi@[#1]#2{\def\@tempa{#1}\ifx\@tempa\@empty \href
  {http://dx.doi.org/#2} {doi:#2}\else \href {http://dx.doi.org/#2} {#1}\fi
  \endgroup}
\def\mn@eprint#1#2{\mn@eprint@#1:#2::\@nil}
\def\mn@eprint@arXiv#1{\href {http://arxiv.org/abs/#1} {{\tt arXiv:#1}}}
\def\mn@eprint@dblp#1{\href {http://dblp.uni-trier.de/rec/bibtex/#1.xml}
  {dblp:#1}}
\def\mn@eprint@#1:#2:#3:#4\@nil{\def\@tempa {#1}\def\@tempb {#2}\def\@tempc
  {#3}\ifx \@tempc \@empty \let \@tempc \@tempb \let \@tempb \@tempa \fi \ifx
  \@tempb \@empty \def\@tempb {arXiv}\fi \@ifundefined
  {mn@eprint@\@tempb}{\@tempb:\@tempc}{\expandafter \expandafter \csname
  mn@eprint@\@tempb\endcsname \expandafter{\@tempc}}}

\bibitem[\protect\citeauthoryear{{Abbott} et~al.,}{{Abbott}
  et~al.}{2016a}]{Abbott16a}
{Abbott} B.~P.,  et~al., 2016a, \mn@doi [Physical Review Letters]
  {10.1103/PhysRevLett.116.061102}, \href
  {http://adsabs.harvard.edu/abs/2016PhRvL.116f1102A} {116, 061102}

\bibitem[\protect\citeauthoryear{{Abbott} et~al.,}{{Abbott}
  et~al.}{2016b}]{Abbott16b}
{Abbott} B.~P.,  et~al., 2016b, \mn@doi [Physical Review Letters]
  {10.1103/PhysRevLett.116.241103}, \href
  {http://adsabs.harvard.edu/abs/2016PhRvL.116x1103A} {116, 241103}

\bibitem[\protect\citeauthoryear{{Abbott} et~al.,}{{Abbott}
  et~al.}{2017a}]{Abbott17a}
{Abbott} B.~P.,  et~al., 2017a, \mn@doi [Physical Review Letters]
  {10.1103/PhysRevLett.118.221101}, \href
  {http://adsabs.harvard.edu/abs/2017PhRvL.118v1101A} {118, 221101}

\bibitem[\protect\citeauthoryear{{Abbott} et~al.,}{{Abbott}
  et~al.}{2017b}]{Abbott17c}
{Abbott} B.~P.,  et~al., 2017b, \mn@doi [Physical Review Letters]
  {10.1103/PhysRevLett.119.141101}, \href
  {http://adsabs.harvard.edu/abs/2017PhRvL.119n1101A} {119, 141101}

\bibitem[\protect\citeauthoryear{{Abbott} et~al.,}{{Abbott}
  et~al.}{2017c}]{Abbott17b}
{Abbott} B.~P.,  et~al., 2017c, \mn@doi [\apjl] {10.3847/2041-8213/aa9f0c},
  \href {http://adsabs.harvard.edu/abs/2017ApJ...851L..35A} {851, L35}

\bibitem[\protect\citeauthoryear{{Agarwal} \& {Milosavljevi{\'c}}}{{Agarwal} \&
  {Milosavljevi{\'c}}}{2011}]{Agarwal11}
{Agarwal} M.,  {Milosavljevi{\'c}} M.,  2011, \mn@doi [\apj]
  {10.1088/0004-637X/729/1/35}, \href
  {http://adsabs.harvard.edu/abs/2011ApJ...729...35A} {729, 35}

\bibitem[\protect\citeauthoryear{{Agarwal}, {Khochfar}, {Johnson}, {Neistein},
  {Dalla Vecchia}  \& {Livio}}{{Agarwal} et~al.}{2012}]{Agarwal12}
{Agarwal} B.,  {Khochfar} S.,  {Johnson} J.~L.,  {Neistein} E.,  {Dalla
  Vecchia} C.,   {Livio} M.,  2012, \mn@doi [\mnras]
  {10.1111/j.1365-2966.2012.21651.x}, \href
  {http://adsabs.harvard.edu/abs/2012MNRAS.425.2854A} {425, 2854}

\bibitem[\protect\citeauthoryear{{Anderson} \& {van der Marel}}{{Anderson} \&
  {van der Marel}}{2010}]{Anderson10}
{Anderson} J.,  {van der Marel} R.~P.,  2010, \mn@doi [\apj]
  {10.1088/0004-637X/710/2/1032}, \href
  {http://adsabs.harvard.edu/abs/2010ApJ...710.1032A} {710, 1032}

\bibitem[\protect\citeauthoryear{{Antonini}, {Capuzzo-Dolcetta},
  {Mastrobuono-Battisti}  \& {Merritt}}{{Antonini} et~al.}{2012}]{Antonini12}
{Antonini} F.,  {Capuzzo-Dolcetta} R.,  {Mastrobuono-Battisti} A.,   {Merritt}
  D.,  2012, \mn@doi [\apj] {10.1088/0004-637X/750/2/111}, \href
  {http://adsabs.harvard.edu/abs/2012ApJ...750..111A} {750, 111}

\bibitem[\protect\citeauthoryear{{Askar}, {Bianchini}, {de Vita}, {Giersz},
  {Hypki}  \& {Kamann}}{{Askar} et~al.}{2017}]{Askar17}
{Askar} A.,  {Bianchini} P.,  {de Vita} R.,  {Giersz} M.,  {Hypki} A.,
  {Kamann} S.,  2017, \mn@doi [\mnras] {10.1093/mnras/stw2573}, \href
  {http://adsabs.harvard.edu/abs/2017MNRAS.464.3090A} {464, 3090}

\bibitem[\protect\citeauthoryear{{Baumgardt}, {Hut}, {Makino}, {McMillan}  \&
  {Portegies Zwart}}{{Baumgardt} et~al.}{2003}]{Baumgardt03}
{Baumgardt} H.,  {Hut} P.,  {Makino} J.,  {McMillan} S.,   {Portegies Zwart}
  S.,  2003, \mn@doi [\apjl] {10.1086/367537}, \href
  {http://adsabs.harvard.edu/abs/2003ApJ...582L..21B} {582, L21}

\bibitem[\protect\citeauthoryear{{Begelman}, {Volonteri}  \& {Rees}}{{Begelman}
  et~al.}{2006}]{Begelman06}
{Begelman} M.~C.,  {Volonteri} M.,   {Rees} M.~J.,  2006, \mn@doi [\mnras]
  {10.1111/j.1365-2966.2006.10467.x}, \href
  {http://adsabs.harvard.edu/abs/2006MNRAS.370..289B} {370, 289}

\bibitem[\protect\citeauthoryear{{Bland-Hawthorn} \&
  {Gerhard}}{{Bland-Hawthorn} \& {Gerhard}}{2016}]{BlandHawthorn16}
{Bland-Hawthorn} J.,  {Gerhard} O.,  2016, \mn@doi [\araa]
  {10.1146/annurev-astro-081915-023441}, \href
  {http://adsabs.harvard.edu/abs/2016ARA%26A..54..529B} {54, 529}

\bibitem[\protect\citeauthoryear{{Bleuler} \& {Teyssier}}{{Bleuler} \&
  {Teyssier}}{2014}]{Bleuler14}
{Bleuler} A.,  {Teyssier} R.,  2014, \mn@doi [\mnras] {10.1093/mnras/stu2005},
  \href {http://adsabs.harvard.edu/abs/2014MNRAS.445.4015B} {445, 4015}

\bibitem[\protect\citeauthoryear{{Boehle} et~al.,}{{Boehle}
  et~al.}{2016}]{Boehle16}
{Boehle} A.,  et~al., 2016, \mn@doi [\apj] {10.3847/0004-637X/830/1/17}, \href
  {http://adsabs.harvard.edu/abs/2016ApJ...830...17B} {830, 17}

\bibitem[\protect\citeauthoryear{{Bromm} \& {Loeb}}{{Bromm} \&
  {Loeb}}{2003}]{Bromm03}
{Bromm} V.,  {Loeb} A.,  2003, \mn@doi [\apj] {10.1086/377529}, \href
  {http://adsabs.harvard.edu/abs/2003ApJ...596...34B} {596, 34}

\bibitem[\protect\citeauthoryear{{Capuzzo-Dolcetta}}{{Capuzzo-Dolcetta}}{1993}]{Capuzzo93}
{Capuzzo-Dolcetta} R.,  1993, \mn@doi [\apj] {10.1086/173189}, \href
  {http://adsabs.harvard.edu/abs/1993ApJ...415..616C} {415, 616}

\bibitem[\protect\citeauthoryear{{Capuzzo-Dolcetta} \&
  {Miocchi}}{{Capuzzo-Dolcetta} \& {Miocchi}}{2008}]{Capuzzo08}
{Capuzzo-Dolcetta} R.,  {Miocchi} P.,  2008, \mn@doi [\apj] {10.1086/588017},
  \href {http://adsabs.harvard.edu/abs/2008ApJ...681.1136C} {681, 1136}

\bibitem[\protect\citeauthoryear{{Casella}, {Ponti}, {Patruno}, {Belloni},
  {Miniutti}  \& {Zampieri}}{{Casella} et~al.}{2008}]{Casella08}
{Casella} P.,  {Ponti} G.,  {Patruno} A.,  {Belloni} T.,  {Miniutti} G.,
  {Zampieri} L.,  2008, \mn@doi [\mnras] {10.1111/j.1365-2966.2008.13372.x},
  \href {http://adsabs.harvard.edu/abs/2008MNRAS.387.1707C} {387, 1707}

\bibitem[\protect\citeauthoryear{{Colgate}}{{Colgate}}{1967}]{Colgate67}
{Colgate} S.~A.,  1967, \mn@doi [\apj] {10.1086/149319}, \href
  {http://adsabs.harvard.edu/abs/1967ApJ...150..163C} {150, 163}

\bibitem[\protect\citeauthoryear{{Cotera}, {Erickson}, {Colgan}, {Simpson},
  {Allen}  \& {Burton}}{{Cotera} et~al.}{1996}]{Cotera96}
{Cotera} A.~S.,  {Erickson} E.~F.,  {Colgan} S.~W.~J.,  {Simpson} J.~P.,
  {Allen} D.~A.,   {Burton} M.~G.,  1996, \mn@doi [\apj] {10.1086/177099},
  \href {http://adsabs.harvard.edu/abs/1996ApJ...461..750C} {461, 750}

\bibitem[\protect\citeauthoryear{{Diemand}, {Madau}  \& {Moore}}{{Diemand}
  et~al.}{2005}]{Diemand05}
{Diemand} J.,  {Madau} P.,   {Moore} B.,  2005, \mn@doi [\mnras]
  {10.1111/j.1365-2966.2005.09604.x}, \href
  {http://adsabs.harvard.edu/abs/2005MNRAS.364..367D} {364, 367}

\bibitem[\protect\citeauthoryear{{Ebisuzaki} et~al.,}{{Ebisuzaki}
  et~al.}{2001}]{Ebisuzaki01}
{Ebisuzaki} T.,  et~al., 2001, \mn@doi [\apjl] {10.1086/338118}, \href
  {http://adsabs.harvard.edu/abs/2001ApJ...562L..19E} {562, L19}

\bibitem[\protect\citeauthoryear{{Farrell}, {Webb}, {Barret}, {Godet}  \&
  {Rodrigues}}{{Farrell} et~al.}{2009}]{Farrell09}
{Farrell} S.~A.,  {Webb} N.~A.,  {Barret} D.,  {Godet} O.,   {Rodrigues} J.~M.,
   2009, \mn@doi [\nat] {10.1038/nature08083}, \href
  {http://adsabs.harvard.edu/abs/2009Natur.460...73F} {460, 73}

\bibitem[\protect\citeauthoryear{{Feldmeier-Krause}, {Zhu}, {Neumayer}, {van de
  Ven}, {de Zeeuw}  \& {Sch{\"o}del}}{{Feldmeier-Krause}
  et~al.}{2017}]{FeldmeierKrause17}
{Feldmeier-Krause} A.,  {Zhu} L.,  {Neumayer} N.,  {van de Ven} G.,  {de Zeeuw}
  P.~T.,   {Sch{\"o}del} R.,  2017, \mn@doi [\mnras] {10.1093/mnras/stw3377},
  \href {http://adsabs.harvard.edu/abs/2017MNRAS.466.4040F} {466, 4040}

\bibitem[\protect\citeauthoryear{{Feldmeier} et~al.,}{{Feldmeier}
  et~al.}{2013}]{FeldmeierKrause13}
{Feldmeier} A.,  et~al., 2013, \mn@doi [\aap] {10.1051/0004-6361/201321168},
  \href {http://adsabs.harvard.edu/abs/2013A%26A...554A..63F} {554, A63}

\bibitem[\protect\citeauthoryear{{Feng} \& {Kaaret}}{{Feng} \&
  {Kaaret}}{2007}]{Feng07}
{Feng} H.,  {Kaaret} P.,  2007, \mn@doi [\apjl] {10.1086/518309}, \href
  {http://adsabs.harvard.edu/abs/2007ApJ...660L.113F} {660, L113}

\bibitem[\protect\citeauthoryear{{Feng} \& {Soria}}{{Feng} \&
  {Soria}}{2011}]{Feng11}
{Feng} H.,  {Soria} R.,  2011, \mn@doi [\nar] {10.1016/j.newar.2011.08.002},
  \href {http://adsabs.harvard.edu/abs/2011NewAR..55..166F} {55, 166}

\bibitem[\protect\citeauthoryear{{Ferri{\`e}re}}{{Ferri{\`e}re}}{2009}]{Ferriere09}
{Ferri{\`e}re} K.,  2009, \mn@doi [\aap] {10.1051/0004-6361/200912617}, \href
  {http://adsabs.harvard.edu/abs/2009A%26A...505.1183F} {505, 1183}

\bibitem[\protect\citeauthoryear{{Figer}, {McLean}  \& {Morris}}{{Figer}
  et~al.}{1995}]{Figer95}
{Figer} D.~F.,  {McLean} I.~S.,   {Morris} M.,  1995, \mn@doi [\apjl]
  {10.1086/309551}, \href {http://adsabs.harvard.edu/abs/1995ApJ...447L..29F}
  {447, L29}

\bibitem[\protect\citeauthoryear{{Fragione}, {Ginsburg}  \&
  {Kocsis}}{{Fragione} et~al.}{2018}]{Fragione18}
{Fragione} G.,  {Ginsburg} I.,   {Kocsis} B.,  2018, \mn@doi [\apj]
  {10.3847/1538-4357/aab368}, \href
  {http://adsabs.harvard.edu/abs/2018ApJ...856...92F} {856, 92}

\bibitem[\protect\citeauthoryear{{Freitag}, {G{\"u}rkan}  \& {Rasio}}{{Freitag}
  et~al.}{2006}]{Freitag06}
{Freitag} M.,  {G{\"u}rkan} M.~A.,   {Rasio} F.~A.,  2006, \mn@doi [\mnras]
  {10.1111/j.1365-2966.2006.10096.x}, \href
  {http://adsabs.harvard.edu/abs/2006MNRAS.368..141F} {368, 141}

\bibitem[\protect\citeauthoryear{{Fritz} et~al.,}{{Fritz}
  et~al.}{2010}]{Fritz10}
{Fritz} T.~K.,  et~al., 2010, \mn@doi [\apj] {10.1088/0004-637X/721/1/395},
  \href {http://adsabs.harvard.edu/abs/2010ApJ...721..395F} {721, 395}

\bibitem[\protect\citeauthoryear{{Fritz} et~al.,}{{Fritz}
  et~al.}{2016}]{Fritz16}
{Fritz} T.~K.,  et~al., 2016, \mn@doi [\apj] {10.3847/0004-637X/821/1/44},
  \href {http://adsabs.harvard.edu/abs/2016ApJ...821...44F} {821, 44}

\bibitem[\protect\citeauthoryear{{Fryer}, {Woosley}  \& {Heger}}{{Fryer}
  et~al.}{2001}]{Fryer01}
{Fryer} C.~L.,  {Woosley} S.~E.,   {Heger} A.,  2001, \mn@doi [\apj]
  {10.1086/319719}, \href {http://adsabs.harvard.edu/abs/2001ApJ...550..372F}
  {550, 372}

\bibitem[\protect\citeauthoryear{{Gallego-Cano}, {Sch{\"o}del}, {Dong},
  {Nogueras-Lara}, {Gallego-Calvente}, {Amaro-Seoane}  \&
  {Baumgardt}}{{Gallego-Cano} et~al.}{2018}]{GallegoCano18}
{Gallego-Cano} E.,  {Sch{\"o}del} R.,  {Dong} H.,  {Nogueras-Lara} F.,
  {Gallego-Calvente} A.~T.,  {Amaro-Seoane} P.,   {Baumgardt} H.,  2018,
  \mn@doi [\aap] {10.1051/0004-6361/201730451}, \href
  {http://adsabs.harvard.edu/abs/2018A%26A...609A..26G} {609, A26}

\bibitem[\protect\citeauthoryear{{Gebhardt}, {Rich}  \& {Ho}}{{Gebhardt}
  et~al.}{2002}]{Gebhardt02}
{Gebhardt} K.,  {Rich} R.~M.,   {Ho} L.~C.,  2002, \mn@doi [\apjl]
  {10.1086/342980}, \href {http://adsabs.harvard.edu/abs/2002ApJ...578L..41G}
  {578, L41}

\bibitem[\protect\citeauthoryear{{Gebhardt}, {Rich}  \& {Ho}}{{Gebhardt}
  et~al.}{2005}]{Gebhardt05}
{Gebhardt} K.,  {Rich} R.~M.,   {Ho} L.~C.,  2005, \mn@doi [\apj]
  {10.1086/497023}, \href {http://adsabs.harvard.edu/abs/2005ApJ...634.1093G}
  {634, 1093}

\bibitem[\protect\citeauthoryear{{Gieles}, {Balbinot}, {Yaaqib},
  {H{\'e}nault-Brunet}, {Zocchi}, {Peuten}  \& {Jonker}}{{Gieles}
  et~al.}{2018}]{Gieles18}
{Gieles} M.,  {Balbinot} E.,  {Yaaqib} R.~I.~S.~M.,  {H{\'e}nault-Brunet} V.,
  {Zocchi} A.,  {Peuten} M.,   {Jonker} P.~G.,  2018, \mn@doi [\mnras]
  {10.1093/mnras/stx2694}, \href
  {http://adsabs.harvard.edu/abs/2018MNRAS.473.4832G} {473, 4832}

\bibitem[\protect\citeauthoryear{{Giersz}, {Leigh}, {Hypki}, {L{\"u}tzgendorf}
  \& {Askar}}{{Giersz} et~al.}{2015}]{Giersz15}
{Giersz} M.,  {Leigh} N.,  {Hypki} A.,  {L{\"u}tzgendorf} N.,   {Askar} A.,
  2015, \mn@doi [\mnras] {10.1093/mnras/stv2162}, \href
  {http://adsabs.harvard.edu/abs/2015MNRAS.454.3150G} {454, 3150}

\bibitem[\protect\citeauthoryear{{Gillessen} et~al.,}{{Gillessen}
  et~al.}{2017}]{Gillessen17}
{Gillessen} S.,  et~al., 2017, \mn@doi [\apj] {10.3847/1538-4357/aa5c41}, \href
  {http://adsabs.harvard.edu/abs/2017ApJ...837...30G} {837, 30}

\bibitem[\protect\citeauthoryear{{Gladstone}, {Roberts}  \& {Done}}{{Gladstone}
  et~al.}{2009}]{Gladstone09}
{Gladstone} J.~C.,  {Roberts} T.~P.,   {Done} C.,  2009, \mn@doi [\mnras]
  {10.1111/j.1365-2966.2009.15123.x}, \href
  {http://adsabs.harvard.edu/abs/2009MNRAS.397.1836G} {397, 1836}

\bibitem[\protect\citeauthoryear{{Gnedin}, {Ostriker}  \& {Tremaine}}{{Gnedin}
  et~al.}{2014}]{Gnedin14}
{Gnedin} O.~Y.,  {Ostriker} J.~P.,   {Tremaine} S.,  2014, \mn@doi [\apj]
  {10.1088/0004-637X/785/1/71}, \href
  {http://adsabs.harvard.edu/abs/2014ApJ...785...71G} {785, 71}

\bibitem[\protect\citeauthoryear{{Goad}, {Roberts}, {Reeves}  \&
  {Uttley}}{{Goad} et~al.}{2006}]{Goad06}
{Goad} M.~R.,  {Roberts} T.~P.,  {Reeves} J.~N.,   {Uttley} P.,  2006, \mn@doi
  [\mnras] {10.1111/j.1365-2966.2005.09702.x}, \href
  {http://adsabs.harvard.edu/abs/2006MNRAS.365..191G} {365, 191}

\bibitem[\protect\citeauthoryear{{Godet}, {Barret}, {Webb}, {Farrell}  \&
  {Gehrels}}{{Godet} et~al.}{2009}]{Godet09}
{Godet} O.,  {Barret} D.,  {Webb} N.~A.,  {Farrell} S.~A.,   {Gehrels} N.,
  2009, \mn@doi [\apjl] {10.1088/0004-637X/705/2/L109}, \href
  {http://adsabs.harvard.edu/abs/2009ApJ...705L.109G} {705, L109}

\bibitem[\protect\citeauthoryear{{Ibata} et~al.,}{{Ibata}
  et~al.}{2009}]{Ibata09}
{Ibata} R.,  et~al., 2009, \mn@doi [\apjl] {10.1088/0004-637X/699/2/L169},
  \href {http://adsabs.harvard.edu/abs/2009ApJ...699L.169I} {699, L169}

\bibitem[\protect\citeauthoryear{{King} \& {Lasota}}{{King} \&
  {Lasota}}{2014}]{King14}
{King} A.,  {Lasota} J.-P.,  2014, \mn@doi [\mnras] {10.1093/mnrasl/slu105},
  \href {http://adsabs.harvard.edu/abs/2014MNRAS.444L..30K} {444, L30}

\bibitem[\protect\citeauthoryear{{K{\i}z{\i}ltan}, {Baumgardt}  \&
  {Loeb}}{{K{\i}z{\i}ltan} et~al.}{2017}]{Kiziltan17}
{K{\i}z{\i}ltan} B.,  {Baumgardt} H.,   {Loeb} A.,  2017, \mn@doi [\nat]
  {10.1038/nature21361}, \href
  {http://adsabs.harvard.edu/abs/2017Natur.542..203K} {542, 203}

\bibitem[\protect\citeauthoryear{{Kormendy} \& {Ho}}{{Kormendy} \&
  {Ho}}{2013}]{Kormendy13}
{Kormendy} J.,  {Ho} L.~C.,  2013, \mn@doi [\araa]
  {10.1146/annurev-astro-082708-101811}, \href
  {http://adsabs.harvard.edu/abs/2013ARA%26A..51..511K} {51, 511}

\bibitem[\protect\citeauthoryear{{Koyama} \& {Inutsuka}}{{Koyama} \&
  {Inutsuka}}{2000}]{Koyama00}
{Koyama} H.,  {Inutsuka} S.-I.,  2000, \mn@doi [\apj] {10.1086/308594}, \href
  {http://adsabs.harvard.edu/abs/2000ApJ...532..980K} {532, 980}

\bibitem[\protect\citeauthoryear{{Lanzoni} et~al.,}{{Lanzoni}
  et~al.}{2013}]{Lanzoni13}
{Lanzoni} B.,  et~al., 2013, \mn@doi [\apj] {10.1088/0004-637X/769/2/107},
  \href {http://adsabs.harvard.edu/abs/2013ApJ...769..107L} {769, 107}

\bibitem[\protect\citeauthoryear{{Larson}}{{Larson}}{1981}]{Larson81}
{Larson} R.~B.,  1981, \mn@doi [\mnras] {10.1093/mnras/194.4.809}, \href
  {http://adsabs.harvard.edu/abs/1981MNRAS.194..809L} {194, 809}

\bibitem[\protect\citeauthoryear{{Larson}}{{Larson}}{1985}]{Larson85}
{Larson} R.~B.,  1985, \mn@doi [\mnras] {10.1093/mnras/214.3.379}, \href
  {http://adsabs.harvard.edu/abs/1985MNRAS.214..379L} {214, 379}

\bibitem[\protect\citeauthoryear{{Larson}}{{Larson}}{2005}]{Larson05}
{Larson} R.~B.,  2005, \mn@doi [\mnras] {10.1111/j.1365-2966.2005.08881.x},
  \href {http://adsabs.harvard.edu/abs/2005MNRAS.359..211L} {359, 211}

\bibitem[\protect\citeauthoryear{{Launhardt}, {Zylka}  \& {Mezger}}{{Launhardt}
  et~al.}{2002}]{Launhardt02}
{Launhardt} R.,  {Zylka} R.,   {Mezger} P.~G.,  2002, \mn@doi [\aap]
  {10.1051/0004-6361:20020017}, \href
  {http://adsabs.harvard.edu/abs/2002A%26A...384..112L} {384, 112}

\bibitem[\protect\citeauthoryear{{Liu}, {Bregman}, {Bai}, {Justham}  \&
  {Crowther}}{{Liu} et~al.}{2013}]{Liu13}
{Liu} J.-F.,  {Bregman} J.~N.,  {Bai} Y.,  {Justham} S.,   {Crowther} P.,
  2013, \mn@doi [\nat] {10.1038/nature12762}, \href
  {http://adsabs.harvard.edu/abs/2013Natur.503..500L} {503, 500}

\bibitem[\protect\citeauthoryear{{Lodato} \& {Natarajan}}{{Lodato} \&
  {Natarajan}}{2007}]{Lodato07}
{Lodato} G.,  {Natarajan} P.,  2007, \mn@doi [\mnras]
  {10.1111/j.1745-3933.2007.00304.x}, \href
  {http://adsabs.harvard.edu/abs/2007MNRAS.377L..64L} {377, L64}

\bibitem[\protect\citeauthoryear{{L{\"u}tzgendorf}, {Kissler-Patig}, {Noyola},
  {Jalali}, {de Zeeuw}, {Gebhardt}  \& {Baumgardt}}{{L{\"u}tzgendorf}
  et~al.}{2011}]{Lutzgendorf11}
{L{\"u}tzgendorf} N.,  {Kissler-Patig} M.,  {Noyola} E.,  {Jalali} B.,  {de
  Zeeuw} P.~T.,  {Gebhardt} K.,   {Baumgardt} H.,  2011, \mn@doi [\aap]
  {10.1051/0004-6361/201116618}, \href
  {http://adsabs.harvard.edu/abs/2011A%26A...533A..36L} {533, A36}

\bibitem[\protect\citeauthoryear{{L{\"u}tzgendorf} et~al.,}{{L{\"u}tzgendorf}
  et~al.}{2013}]{Lutzgendorf13}
{L{\"u}tzgendorf} N.,  et~al., 2013, \mn@doi [\aap]
  {10.1051/0004-6361/201220307}, \href
  {http://adsabs.harvard.edu/abs/2013A%26A...552A..49L} {552, A49}

\bibitem[\protect\citeauthoryear{{L{\"u}tzgendorf}, {Gebhardt}, {Baumgardt},
  {Noyola}, {Neumayer}, {Kissler-Patig}  \& {de Zeeuw}}{{L{\"u}tzgendorf}
  et~al.}{2015}]{Lutzgendorf15}
{L{\"u}tzgendorf} N.,  {Gebhardt} K.,  {Baumgardt} H.,  {Noyola} E.,
  {Neumayer} N.,  {Kissler-Patig} M.,   {de Zeeuw} T.,  2015, \mn@doi [\aap]
  {10.1051/0004-6361/201425524}, \href
  {http://adsabs.harvard.edu/abs/2015A%26A...581A...1L} {581, A1}

\bibitem[\protect\citeauthoryear{{Madau} \& {Rees}}{{Madau} \&
  {Rees}}{2001}]{Madau01}
{Madau} P.,  {Rees} M.~J.,  2001, \mn@doi [\apjl] {10.1086/319848}, \href
  {http://adsabs.harvard.edu/abs/2001ApJ...551L..27M} {551, L27}

\bibitem[\protect\citeauthoryear{{Mapelli}}{{Mapelli}}{2016}]{Mapelli16}
{Mapelli} M.,  2016, \mn@doi [\mnras] {10.1093/mnras/stw869}, \href
  {http://adsabs.harvard.edu/abs/2016MNRAS.459.3432M} {459, 3432}

\bibitem[\protect\citeauthoryear{{Mapelli}, {Colpi}  \& {Zampieri}}{{Mapelli}
  et~al.}{2009}]{Mapelli09}
{Mapelli} M.,  {Colpi} M.,   {Zampieri} L.,  2009, \mn@doi [\mnras]
  {10.1111/j.1745-3933.2009.00645.x}, \href
  {http://adsabs.harvard.edu/abs/2009MNRAS.395L..71M} {395, L71}

\bibitem[\protect\citeauthoryear{{Mapelli}, {Ripamonti}, {Zampieri}, {Colpi}
  \& {Bressan}}{{Mapelli} et~al.}{2010}]{Mapelli10}
{Mapelli} M.,  {Ripamonti} E.,  {Zampieri} L.,  {Colpi} M.,   {Bressan} A.,
  2010, \mn@doi [\mnras] {10.1111/j.1365-2966.2010.17048.x}, \href
  {http://adsabs.harvard.edu/abs/2010MNRAS.408..234M} {408, 234}

\bibitem[\protect\citeauthoryear{{Mastrobuono-Battisti}, {Perets}  \&
  {Loeb}}{{Mastrobuono-Battisti} et~al.}{2014}]{Mastrobuono14}
{Mastrobuono-Battisti} A.,  {Perets} H.~B.,   {Loeb} A.,  2014, \mn@doi [\apj]
  {10.1088/0004-637X/796/1/40}, \href
  {http://adsabs.harvard.edu/abs/2014ApJ...796...40M} {796, 40}

\bibitem[\protect\citeauthoryear{{McKernan}, {Ford}, {Lyra}  \&
  {Perets}}{{McKernan} et~al.}{2012}]{McKernan12}
{McKernan} B.,  {Ford} K.~E.~S.,  {Lyra} W.,   {Perets} H.~B.,  2012, \mn@doi
  [\mnras] {10.1111/j.1365-2966.2012.21486.x}, \href
  {http://adsabs.harvard.edu/abs/2012MNRAS.425..460M} {425, 460}

\bibitem[\protect\citeauthoryear{{McMillan} \& {Portegies Zwart}}{{McMillan} \&
  {Portegies Zwart}}{2003}]{McMillan03}
{McMillan} S.~L.~W.,  {Portegies Zwart} S.~F.,  2003, \mn@doi [\apj]
  {10.1086/377577}, \href {http://adsabs.harvard.edu/abs/2003ApJ...596..314M}
  {596, 314}

\bibitem[\protect\citeauthoryear{{Mezcua}}{{Mezcua}}{2017}]{Mezcua17}
{Mezcua} M.,  2017, \mn@doi [International Journal of Modern Physics D]
  {10.1142/S021827181730021X}, \href
  {http://adsabs.harvard.edu/abs/2017IJMPD..2630021M} {26, 1730021}

\bibitem[\protect\citeauthoryear{{Mezcua}, {Roberts}, {Lobanov}  \&
  {Sutton}}{{Mezcua} et~al.}{2015}]{Mezcua15}
{Mezcua} M.,  {Roberts} T.~P.,  {Lobanov} A.~P.,   {Sutton} A.~D.,  2015,
  \mn@doi [\mnras] {10.1093/mnras/stv143}, \href
  {http://adsabs.harvard.edu/abs/2015MNRAS.448.1893M} {448, 1893}

\bibitem[\protect\citeauthoryear{{Miller} \& {Davies}}{{Miller} \&
  {Davies}}{2012}]{Miller12}
{Miller} M.~C.,  {Davies} M.~B.,  2012, \mn@doi [\apj]
  {10.1088/0004-637X/755/1/81}, \href
  {http://adsabs.harvard.edu/abs/2012ApJ...755...81M} {755, 81}

\bibitem[\protect\citeauthoryear{{Miller} \& {Hamilton}}{{Miller} \&
  {Hamilton}}{2002}]{Miller02}
{Miller} M.~C.,  {Hamilton} D.~P.,  2002, \mn@doi [\mnras]
  {10.1046/j.1365-8711.2002.05112.x}, \href
  {http://adsabs.harvard.edu/abs/2002MNRAS.330..232C} {330, 232}

\bibitem[\protect\citeauthoryear{{Miller}, {Fabbiano}, {Miller}  \&
  {Fabian}}{{Miller} et~al.}{2003}]{Miller03}
{Miller} J.~M.,  {Fabbiano} G.,  {Miller} M.~C.,   {Fabian} A.~C.,  2003,
  \mn@doi [\apjl] {10.1086/368373}, \href
  {http://adsabs.harvard.edu/abs/2003ApJ...585L..37M} {585, L37}

\bibitem[\protect\citeauthoryear{{Miocchi}, {Capuzzo Dolcetta}, {Di Matteo}  \&
  {Vicari}}{{Miocchi} et~al.}{2006}]{Miocchi06}
{Miocchi} P.,  {Capuzzo Dolcetta} R.,  {Di Matteo} P.,   {Vicari} A.,  2006,
  \mn@doi [\apj] {10.1086/503663}, \href
  {http://adsabs.harvard.edu/abs/2006ApJ...644..940M} {644, 940}

\bibitem[\protect\citeauthoryear{{Mouschovias} \& {Spitzer}}{{Mouschovias} \&
  {Spitzer}}{1976}]{Mouschovias76}
{Mouschovias} T.~C.,  {Spitzer} Jr. L.,  1976, \mn@doi [\apj] {10.1086/154835},
  \href {http://adsabs.harvard.edu/abs/1976ApJ...210..326M} {210, 326}

\bibitem[\protect\citeauthoryear{{Nagata}, {Woodward}, {Shure}, {Pipher}  \&
  {Okuda}}{{Nagata} et~al.}{1990}]{Nagata90}
{Nagata} T.,  {Woodward} C.~E.,  {Shure} M.,  {Pipher} J.~L.,   {Okuda} H.,
  1990, \mn@doi [\apj] {10.1086/168446}, \href
  {http://adsabs.harvard.edu/abs/1990ApJ...351...83N} {351, 83}

\bibitem[\protect\citeauthoryear{{Nagata}, {Woodward}, {Shure}  \&
  {Kobayashi}}{{Nagata} et~al.}{1995}]{Nagata95}
{Nagata} T.,  {Woodward} C.~E.,  {Shure} M.,   {Kobayashi} N.,  1995, \mn@doi
  [\aj] {10.1086/117395}, \href
  {http://adsabs.harvard.edu/abs/1995AJ....109.1676N} {109, 1676}

\bibitem[\protect\citeauthoryear{{Noyola}, {Gebhardt}  \& {Bergmann}}{{Noyola}
  et~al.}{2008}]{Noyola08}
{Noyola} E.,  {Gebhardt} K.,   {Bergmann} M.,  2008, \mn@doi [\apj]
  {10.1086/529002}, \href {http://adsabs.harvard.edu/abs/2008ApJ...676.1008N}
  {676, 1008}

\bibitem[\protect\citeauthoryear{{Noyola}, {Gebhardt}, {Kissler-Patig},
  {L{\"u}tzgendorf}, {Jalali}, {de Zeeuw}  \& {Baumgardt}}{{Noyola}
  et~al.}{2010}]{Noyola10}
{Noyola} E.,  {Gebhardt} K.,  {Kissler-Patig} M.,  {L{\"u}tzgendorf} N.,
  {Jalali} B.,  {de Zeeuw} P.~T.,   {Baumgardt} H.,  2010, \mn@doi [\apjl]
  {10.1088/2041-8205/719/1/L60}, \href
  {http://adsabs.harvard.edu/abs/2010ApJ...719L..60N} {719, L60}

\bibitem[\protect\citeauthoryear{{Oka}, {Mizuno}, {Miura}  \& {Takekawa}}{{Oka}
  et~al.}{2016}]{Oka16}
{Oka} T.,  {Mizuno} R.,  {Miura} K.,   {Takekawa} S.,  2016, \mn@doi [\apjl]
  {10.3847/2041-8205/816/1/L7}, \href
  {http://adsabs.harvard.edu/abs/2016ApJ...816L...7O} {816, L7}

\bibitem[\protect\citeauthoryear{{Oka}, {Tsujimoto}, {Iwata}, {Nomura}  \&
  {Takekawa}}{{Oka} et~al.}{2017}]{Oka17}
{Oka} T.,  {Tsujimoto} S.,  {Iwata} Y.,  {Nomura} M.,   {Takekawa} S.,  2017,
  \mn@doi [Nature Astronomy] {10.1038/s41550-017-0224-z}, \href
  {http://adsabs.harvard.edu/abs/2017NatAs...1..709O} {1, 709}

\bibitem[\protect\citeauthoryear{{Perera} et~al.,}{{Perera}
  et~al.}{2017}]{Perera17}
{Perera} B.~B.~P.,  et~al., 2017, \mn@doi [\mnras] {10.1093/mnras/stx501},
  \href {http://adsabs.harvard.edu/abs/2017MNRAS.468.2114P} {468, 2114}

\bibitem[\protect\citeauthoryear{{Portegies Zwart}, {Baumgardt}, {Hut},
  {Makino}  \& {McMillan}}{{Portegies Zwart} et~al.}{2004}]{PortegiesZwart04}
{Portegies Zwart} S.~F.,  {Baumgardt} H.,  {Hut} P.,  {Makino} J.,   {McMillan}
  S.~L.~W.,  2004, \mn@doi [\nat] {10.1038/nature02448}, \href
  {http://adsabs.harvard.edu/abs/2004Natur.428..724P} {428, 724}

\bibitem[\protect\citeauthoryear{{Portegies Zwart}, {Baumgardt}, {McMillan},
  {Makino}, {Hut}  \& {Ebisuzaki}}{{Portegies Zwart}
  et~al.}{2006}]{PortegiesZwart06}
{Portegies Zwart} S.~F.,  {Baumgardt} H.,  {McMillan} S.~L.~W.,  {Makino} J.,
  {Hut} P.,   {Ebisuzaki} T.,  2006, \mn@doi [\apj] {10.1086/500361}, \href
  {http://adsabs.harvard.edu/abs/2006ApJ...641..319P} {641, 319}

\bibitem[\protect\citeauthoryear{{Ravi}, {Vedantham}  \& {Phinney}}{{Ravi}
  et~al.}{2018}]{Ravi18}
{Ravi} V.,  {Vedantham} H.,   {Phinney} E.~S.,  2018, \mn@doi [\mnras]
  {10.1093/mnrasl/sly077}, \href
  {http://adsabs.harvard.edu/abs/2018MNRAS.478L..72R} {478, L72}

\bibitem[\protect\citeauthoryear{{Reines} \& {Volonteri}}{{Reines} \&
  {Volonteri}}{2015}]{Reines15}
{Reines} A.~E.,  {Volonteri} M.,  2015, \mn@doi [\apj]
  {10.1088/0004-637X/813/2/82}, \href
  {http://adsabs.harvard.edu/abs/2015ApJ...813...82R} {813, 82}

\bibitem[\protect\citeauthoryear{{Schneider}, {Ferrara}, {Natarajan}  \&
  {Omukai}}{{Schneider} et~al.}{2002}]{Schneider02}
{Schneider} R.,  {Ferrara} A.,  {Natarajan} P.,   {Omukai} K.,  2002, \mn@doi
  [\apj] {10.1086/339917}, \href
  {http://adsabs.harvard.edu/abs/2002ApJ...571...30S} {571, 30}

\bibitem[\protect\citeauthoryear{{Sch{\"o}del}, {Eckart}, {Iserlohe}, {Genzel}
  \& {Ott}}{{Sch{\"o}del} et~al.}{2005}]{Schodel05}
{Sch{\"o}del} R.,  {Eckart} A.,  {Iserlohe} C.,  {Genzel} R.,   {Ott} T.,
  2005, \mn@doi [\apjl] {10.1086/431307}, \href
  {http://adsabs.harvard.edu/abs/2005ApJ...625L.111S} {625, L111}

\bibitem[\protect\citeauthoryear{{Serabyn}, {Shupe}  \& {Figer}}{{Serabyn}
  et~al.}{1998}]{Serabyn98}
{Serabyn} E.,  {Shupe} D.,   {Figer} D.~F.,  1998, \mn@doi [\nat]
  {10.1038/28799}, \href {http://adsabs.harvard.edu/abs/1998Natur.394..448S}
  {394, 448}

\bibitem[\protect\citeauthoryear{{Shang}, {Bryan}  \& {Haiman}}{{Shang}
  et~al.}{2010}]{Shang10}
{Shang} C.,  {Bryan} G.~L.,   {Haiman} Z.,  2010, \mn@doi [\mnras]
  {10.1111/j.1365-2966.2009.15960.x}, \href
  {http://adsabs.harvard.edu/abs/2010MNRAS.402.1249S} {402, 1249}

\bibitem[\protect\citeauthoryear{{Spera} \& {Mapelli}}{{Spera} \&
  {Mapelli}}{2017}]{Spera17}
{Spera} M.,  {Mapelli} M.,  2017, \mn@doi [\mnras] {10.1093/mnras/stx1576},
  \href {http://adsabs.harvard.edu/abs/2017MNRAS.470.4739S} {470, 4739}

\bibitem[\protect\citeauthoryear{{Stone}, {K{\"u}pper}  \& {Ostriker}}{{Stone}
  et~al.}{2017}]{Stone17}
{Stone} N.~C.,  {K{\"u}pper} A.~H.~W.,   {Ostriker} J.~P.,  2017, \mn@doi
  [\mnras] {10.1093/mnras/stx097}, \href
  {http://adsabs.harvard.edu/abs/2017MNRAS.467.4180S} {467, 4180}

\bibitem[\protect\citeauthoryear{{Sutton}, {Roberts}, {Walton}, {Gladstone}  \&
  {Scott}}{{Sutton} et~al.}{2012}]{Sutton12}
{Sutton} A.~D.,  {Roberts} T.~P.,  {Walton} D.~J.,  {Gladstone} J.~C.,
  {Scott} A.~E.,  2012, \mn@doi [\mnras] {10.1111/j.1365-2966.2012.20944.x},
  \href {http://adsabs.harvard.edu/abs/2012MNRAS.423.1154S} {423, 1154}

\bibitem[\protect\citeauthoryear{{Tanaka}}{{Tanaka}}{2018}]{Tanaka18}
{Tanaka} K.,  2018, preprint, \href
  {http://adsabs.harvard.edu/abs/2018arXiv180403661T} {} (\mn@eprint {arXiv}
  {1804.03661})

\bibitem[\protect\citeauthoryear{{Tanaka}, {Oka}, {Matsumura}, {Nagai}  \&
  {Kamegai}}{{Tanaka} et~al.}{2014}]{Tanaka14}
{Tanaka} K.,  {Oka} T.,  {Matsumura} S.,  {Nagai} M.,   {Kamegai} K.,  2014,
  \mn@doi [\apj] {10.1088/0004-637X/783/1/62}, \href
  {http://adsabs.harvard.edu/abs/2014ApJ...783...62T} {783, 62}

\bibitem[\protect\citeauthoryear{{Teyssier}}{{Teyssier}}{2002}]{Teyssier02}
{Teyssier} R.,  2002, \mn@doi [\aap] {10.1051/0004-6361:20011817}, \href
  {http://adsabs.harvard.edu/abs/2002A%26A...385..337T} {385, 337}

\bibitem[\protect\citeauthoryear{Toro}{Toro}{2009}]{Toro09}
Toro E.,  2009, Riemann Solvers and Numerical Methods for Fluid Dynamics: A
  Practical Introduction.
Springer Berlin Heidelberg

\bibitem[\protect\citeauthoryear{{Tremaine}, {Ostriker}  \&
  {Spitzer}}{{Tremaine} et~al.}{1975}]{Tremaine75}
{Tremaine} S.~D.,  {Ostriker} J.~P.,   {Spitzer} Jr. L.,  1975, \mn@doi [\apj]
  {10.1086/153422}, \href {http://adsabs.harvard.edu/abs/1975ApJ...196..407T}
  {196, 407}

\bibitem[\protect\citeauthoryear{{Tsuboi}, {Kitamura}, {Tsutsumi}, {Uehara},
  {Miyoshi}, {Miyawaki}  \& {Miyazaki}}{{Tsuboi} et~al.}{2017}]{Tsuboi17}
{Tsuboi} M.,  {Kitamura} Y.,  {Tsutsumi} T.,  {Uehara} K.,  {Miyoshi} M.,
  {Miyawaki} R.,   {Miyazaki} A.,  2017, \mn@doi [\apjl]
  {10.3847/2041-8213/aa97d3}, \href
  {http://adsabs.harvard.edu/abs/2017ApJ...850L...5T} {850, L5}

\bibitem[\protect\citeauthoryear{{Turk}, {Smith}, {Oishi}, {Skory}, {Skillman},
  {Abel}  \& {Norman}}{{Turk} et~al.}{2011}]{Turk11}
{Turk} M.~J.,  {Smith} B.~D.,  {Oishi} J.~S.,  {Skory} S.,  {Skillman} S.~W.,
  {Abel} T.,   {Norman} M.~L.,  2011, \mn@doi [The Astrophysical Journal
  Supplement Series] {10.1088/0067-0049/192/1/9}, \href
  {http://adsabs.harvard.edu/abs/2011ApJS..192....9T} {192, 9}

\bibitem[\protect\citeauthoryear{{Volonteri} \& {Perna}}{{Volonteri} \&
  {Perna}}{2005}]{Volonteri05}
{Volonteri} M.,  {Perna} R.,  2005, \mn@doi [\mnras]
  {10.1111/j.1365-2966.2005.08832.x}, \href
  {http://adsabs.harvard.edu/abs/2005MNRAS.358..913V} {358, 913}

\bibitem[\protect\citeauthoryear{{Volonteri}, {Haardt}  \& {Madau}}{{Volonteri}
  et~al.}{2003}]{Volonteri03}
{Volonteri} M.,  {Haardt} F.,   {Madau} P.,  2003, \mn@doi [\apj]
  {10.1086/344675}, \href {http://adsabs.harvard.edu/abs/2003ApJ...582..559V}
  {582, 559}

\bibitem[\protect\citeauthoryear{{Yalinewich} \& {Beniamini}}{{Yalinewich} \&
  {Beniamini}}{2017}]{Yalinewich17}
{Yalinewich} A.,  {Beniamini} P.,  2017, preprint, \href
  {http://adsabs.harvard.edu/abs/2017arXiv170905738Y} {} (\mn@eprint {arXiv}
  {1709.05738})

\bibitem[\protect\citeauthoryear{{Zampieri} \& {Roberts}}{{Zampieri} \&
  {Roberts}}{2009}]{Zampieri09}
{Zampieri} L.,  {Roberts} T.~P.,  2009, \mn@doi [\mnras]
  {10.1111/j.1365-2966.2009.15509.x}, \href
  {http://adsabs.harvard.edu/abs/2009MNRAS.400..677Z} {400, 677}

\bibitem[\protect\citeauthoryear{{Zocchi}, {Gieles}  \&
  {H{\'e}nault-Brunet}}{{Zocchi} et~al.}{2017}]{Zocchi17}
{Zocchi} A.,  {Gieles} M.,   {H{\'e}nault-Brunet} V.,  2017, \mn@doi [\mnras]
  {10.1093/mnras/stx316}, \href
  {http://adsabs.harvard.edu/abs/2017MNRAS.468.4429Z} {468, 4429}

\makeatother
\end{thebibliography}

%% Alternatively you could enter them by hand, like this:
%% This method is tedious and prone to error if you have lots of references
%\begin{thebibliography}{99}
%\end{thebibliography}

%%%%%%%%%%%%%%%%%%%%%%%%%%%%%%%%%%%%%%%%%%%%%%%%%%

%%%%%%%%%%%%%%%%% APPENDICES %%%%%%%%%%%%%%%%%%%%%

\appendix

\section{Impact of $E_{orb}$ on the mass of the IMBH}\label{eorb}

In this Appendix we estimate the impact of a non-zero orbital energy of the radial orbit of the cloud towards the IMBH.

For the $E_{orb}<0$ case, the most extreme configuration is that the cloud had zero velocity at $d_{BH,0}$ equal to the current position of the tail of the cloud. This simply means that Eq. \ref{dvlos} becomes

\begin{equation}
\Delta v_{LOS,new}= \sqrt{2G M_{BH} \sin\theta\left[\frac{1}{D}-\frac{1}{S+D}\right]}\cos\theta.
\end{equation}

Thus the mass limit would become

\begin{equation}
M_{BH,new}=M_{BH}(E_{orb}=0)\frac{[1-\sqrt{D/(S+D)}]^2}{1-D/(S+D)}
\end{equation}

For the case $(D,S)$ = (2,14) arcsec, the new lower limit is 0.5 times the value we get for the $E_{orb}=0$ case.

In general, under the assumption of $d_{BH,h}<d_{BH,t}<<d_{BH,0}$, a 1st-order Taylor expansion of Eq. \ref{vff} gives

\begin{equation}
M_{BH,new}\approx M_{BH}(E_{orb}=0)\left(1+\frac{d_{BH,t}-d_{BH,h}}{2d_0}\right)^{-2}.
\end{equation}

The case of $E_{orb}>0$ does not affect our main conclusions, since we provided lower limits on the mass of CO-0.40-0.22*, but this case always provides higher IMBH masses.

%t can be placed in an Appendix which appears after the list of references.

%%%%%%%%%%%%%%%%%%%%%%%%%%%%%%%%%%%%%%%%%%%%%%%%%%

% Don't change these lines
\bsp	% typesetting comment
\label{lastpage}
\end{document}